\documentclass[reprint,aps,prd,notitlepage,superscriptaddress,nofootinbib]{revtex4-2}

\usepackage[utf8]{inputenc}
\usepackage[english]{babel}
\usepackage{amsmath}
\usepackage{amsfonts}
\usepackage{amssymb}
\usepackage{listings}
\usepackage{lipsum}
\usepackage{multirow}
\usepackage{datetime}
\usepackage{graphicx}
\usepackage{mathtools}
\usepackage{mathrsfs}
\usepackage{dcolumn}
\usepackage{multirow}
\usepackage[caption=false]{subfig}
\usepackage{soul}
\usepackage{breqn}
\usepackage{chngcntr}
\usepackage{ulem}
\usepackage{orcidlink}

\usepackage{color} 
\usepackage[dvipsnames]{xcolor}
\usepackage{hyperref}
\hypersetup{
    colorlinks=true, 
    pdfborder = {0 0 0.5 [3 3]},
    anchorcolor=black,
    citecolor=blue,
    linktoc=all,    
    linktocpage=true,
    linkcolor=red,
	urlcolor=blue
}

\newcommand{\centra}{CENTRA, Departamento de Física, Instituto Superior Técnico – IST,
Universidade de Lisboa – UL, Avenida Rovisco Pais 1, 1049-001 Lisboa, Portugal}

\begin{document}

\title{Spectral instabilities in the time domain}

\author{Taillte May\orcidlink{0000-0002-4237-3134}}
\email{taillte.may@tecnico.ulisboa.pt}
\affiliation{\centra}
\affiliation{Centre of Gravity\char`,{} University of Copenhagen\char`,{} Denmark}

\author{Adrien Kuntz\orcidlink{0000-0002-4803-2998}}
\email{adrien.kuntz@tecnico.ulisboa.pt}
\affiliation{\centra}

\author{Nicola Franchini\orcidlink{0000-0002-9939-733X}}
\email{nicola.franchini@tecnico.ulisboa.pt}
\affiliation{\centra}

\author{Valentin Boyanov}
\affiliation{\centra}

\author{Vitor Cardoso}
\affiliation{Centre of Gravity\char`,{} University of Copenhagen\char`,{} Denmark}
\affiliation{\centra}


\begin{abstract} 
We investigate the effect of a spectral instability on the time domain waveform using a one-dimensional double P\"oschl--Teller model. By analytically following successive scatterings between the primary potential barrier and a weak, spatially separated perturbation, we identify a secular contribution that first appears after one causal round trip between the barriers. Before the first echo, all overtones remain at their unperturbed frequencies. After the first echo reaches the observer, there is a secular, linear-in-time correction. When this correction is perturbative, it can be interpreted as a shift of the quasi-normal mode frequency. In that case, we show that its coefficient reproduces the frequency-domain result. More generally, it generates finite-time effective frequencies that need not coincide with either the unperturbed or fully perturbed quasi-normal mode spectrum. We confirm the results of our analytic calculation using numerical time-domain evolutions.
\end{abstract}

\maketitle

\section{Introduction}
Quasinormal modes (QNMs) are the characteristic oscillations of dissipative systems and are described by complex frequencies whose real and imaginary parts determine the oscillation and decay rates, respectively. In General Relativity, the QNM spectrum of an isolated Kerr black hole is fixed entirely by its mass and angular momentum. Measuring multiple ringdown frequencies therefore provides a basis for black-hole spectroscopy, tests of the Kerr hypothesis, and searches for deviations from General Relativity~\cite{Berti:2009kk,Berti:2025hly}.

The QNM spectrum of black holes is unstable. In other words, a perturbation that is small in amplitude can produce a large displacement of the QNM frequencies~\cite{Nollert:1996rf,Barausse:2014tra,Jaramillo:2020tuu,Cardoso:2024mrw,Cardoso:2025npr}. This property was first studied in the context of ad hoc perturbations to the effective potential governing massless fields: a weak\footnote{Here by ``weak'' we mean small in amplitude, rather than necessarily small in energy~\cite{Gasperin:2021kfv}. We also note that a recent study~\cite{DellaRocca:2026zym} suggests that such bumps are difficult to realize with the physical matter distributions considered there.}, spatially-separated deformation of the effective potential can continuously shift the modes from their original black-hole values, change which mode is the least damped, and introduce an additional sequence of long-lived modes~\cite{Cheung:2021bol,Cardoso:2024mrw,Boyanov:2024fgc}. Related spectral changes have also been found in models with environmental matter fields or altered boundary conditions~\cite{Nollert:1996rf,Barausse:2014tra,Jaramillo:2020tuu,Cardoso:2024mrw,Cardoso:2025npr,Cardoso:2022whc}.
We refer collectively to these phenomena as the spectral instability.

Taken at face value, the impact of the spectral instability of black hole QNMs is considerable. A cornerstone of the black hole spectroscopy program is that QNM frequencies of vacuum black holes describe the gravitational-wave ringdown. If any astrophysical environment (or any effect leading to spatially localised perturbations~\cite{Jaramillo:2020tuu,Warnick:2024usx}) changes the QNM spectrum considerably, this would appear to challenge the reliability of black-hole spectroscopy.

It turns out that the sensitivity of the spectrum need not manifest immediately in the time-domain waveform. The QNM spectrum is a global property of the frequency-domain boundary-value problem and therefore responds immediately to a modification of the potential, irrespective of where that modification is located. By contrast, the retarded waveform can only depend on a perturbation once radiation has interacted with it and reached the observer. The additional return to the primary barrier introduces a further delay. Numerical and analytic studies have shown that the prompt ringdown can remain close to the unperturbed signal even when the corresponding QNM spectrum is substantially modified~\cite{Nollert:1996rf,Barausse:2014tra,Berti:2022xfj,Cardoso:2022whc,Yang:2024vor,Ianniccari:2024ysv}. In particular, Ref.~\cite{Yang:2024vor} showed explicitly that the contributions carrying information about the perturbed QNM spectrum are delayed by the additional propagation time associated with scattering from the perturbation. What remains less clear is how this delayed response is related to the shifted spectrum in the time domain, and how the waveform should be characterized at finite times.

We address this question using a one-dimensional double P\"oschl--Teller model, by following the successive scatterings between the primary potential barrier and a weak, spatially separated perturbation. Recent works obtained an analytical solution to the time-domain scattering problem for the Pöschl--Teller potential using the Green's function~\cite{Kuntz:2025gdq,Arnaudo:2025uos}. Here, we build on these results to derive an analytical description of spectral instabilities directly in the time domain. We treat the secondary barrier as a controlled model of a localized perturbation, without assuming that every such profile can be generated by a physical matter distribution.

\begin{figure}[h]
\centering
\includegraphics[width=8cm]{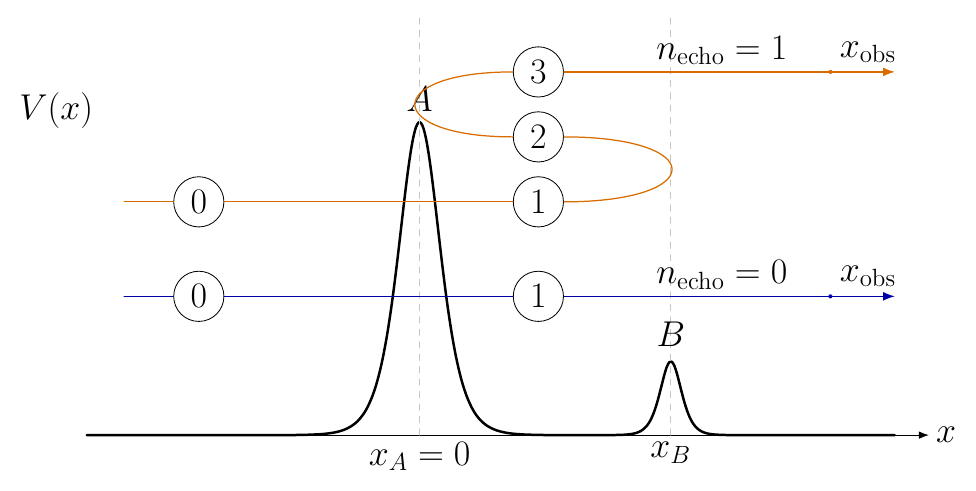}
\caption{This diagram shows the paths for a signal to the left of the potential to reach an observer to the right of the potential. Here we show the paths with $n_{\rm echo}=0,1$.
}
\label{fig:diagram_echo}
\end{figure}

Within the separated-barrier approximation, we show that before the first reflected signal returns (the $n_{\rm echo}=1$ signal in Fig.~\ref{fig:diagram_echo}), none of the original QNM frequencies are modified in the waveform. This statement holds mode by mode, including all overtones. The distant perturbation can modify amplitudes and introduce its own characteristic modes, but it does not shift the original QNM frequencies before the first echo.

Once radiation has completed one causal round trip between the two barriers, we find an additional resonant contribution. This resonant contribution has a secular prefactor, growing linearly with time. In the small-frequency-shift regime, this term can be interpreted as a perturbation of the original QNM frequency, and its coefficient reproduces the analytic frequency-domain result of Ref.~\cite{Ianniccari:2024ysv}.

More generally, however, the finite-time response after a fixed number of scatterings need not be characterized by either the unperturbed or the full perturbed QNM spectrum. Beyond the small-frequency-shift regime, the secular first-echo contribution instead produces effective finite-time frequencies that depend on the fitting interval and need not coincide with either spectrum. We confirm this causal behaviour with numerical time-domain evolutions and show that the analytic prompt-plus-first-echo waveform accurately describes the observed spectral migration during the first-echo window.

Interestingly, this behaviour closely resembles the QNM resonances that arise near exceptional points~\cite{Motohashi:2024fwt,Cavalcante:2024swt,Lo:2025njp,Yang:2025dbn}. As noted in~\cite{Yang:2025dbn}, in beyond-Kerr spacetimes additional parameters of the background can cause two neighbouring QNMs to approach one another and undergo a resonance. Near resonance, the response acquires the characteristic linear-in-time growth familiar from forced resonant systems. This phenomenon is analogous to the behaviour we find here after the first echo. Here, however, the secular term arises from repeated scattering through the same QNM pole and does not require an exceptional point in the spectrum of the combined system.

This distinction between the global QNM spectrum and the causal waveform is particularly relevant when the perturbation is spatially extended or located far from the black hole. In such cases, a large spectral modification may not be visible during the immediate ringdown. The frequency changes associated with repeated scattering are delayed by the corresponding echo time. Extended environments such as accretion disks or other matter clouds provide a possible physical setting in which this separation of scales may be important.

The remainder of this paper is organized as follows. We introduce the P\"oschl-Teller model and the framework for the analytic time domain calculation in Sec.~\ref{sec:preliminaries}. In Sec.~\ref{sec:time_domain_sol}, we construct the time-domain response by successive scatterings, isolating the terms that alter the observed effective frequency. In Sec.~\ref{sec:confirmation_literature} we compare the resulting small-shift expression with existing analytic frequency-domain calculations, while Secs.~\ref{sec:confirmation_numerics_timedomain} and~\ref{sec:confirmation_numerics_freqdomain} test the causal behaviour and finite-time effective frequency evolution against numerical simulations.

\section{Preliminaries}
\label{sec:preliminaries}

\subsection{The Pöschl-Teller Potential}\label{sec:PT_potential}

We consider the behaviour of the wave equation with a potential:
\begin{equation}
   - \frac{\partial^2 \psi}{\partial t^2} + \frac{\partial^2 \psi}{\partial x^2} - V(x) \psi = 0 \label{eq:RWZeq} \; .
\end{equation}

Particularly, we will focus on the interaction of waves with the peak of the potential, and spatially localised perturbations thereof. For simplicity, we will use the one-dimensional Pöschl-Teller potential, as it has similar peak properties to the Regge-Wheeler potential, but has a simpler asymptotic behaviour (leading to a lack of branch cut in the Green's function) and presents a more analytically tractable problem. The P\"oschl-Teller potential has the following form:
\begin{align}
    V_{PT}(x) = \frac{V_A}{\cosh^2 \alpha_A (x-x_A)},
\end{align}
where we take $x_A=0$ without loss of generality. Here, $V_A = V_{PT}(x_A)$ represents the height of the potential at its maximum and $\alpha_A$ is related to the second derivative of $V_{PT}$ at its maximum, $\alpha_A^2 = - \frac{\mathrm{d}^2 V_{PT}}{\mathrm{d}x^2} \big|_{x_A} /(2V_A)$.

The QNM frequencies $\omega_n$ of this potential are
\begin{align}
    \label{eq:qnm_A}
    \omega_{n}^A &= -i\alpha_A \Big(n + \frac{1}{2}+\lambda_A \Big), \\
    \lambda_A &\equiv \pm\frac{i}{2}   \sqrt{\frac{4 V_A}{\alpha_A^2}-1}, \label{eq:lambda_A}
\end{align}
where $n \in \mathbb{N}$, and we have assumed here that $4 V_A > \alpha_A^2$.  This is true if the P\"oschl-Teller potential is tuned to fit the Regge-Wheeler potential. The prompt part of the time-domain Green's function can also be expanded in exponentials with characteristic frequencies $w_k^A$, which we refer to as \textit{Matsubara terms}~\cite{Kuntz:2025gdq,Arnaudo:2025uos,Arnaudo:2025kit}:
\begin{equation} \label{eq:defMatsubaraFreq}
    w_k^A = -i\alpha_A k\ ,
\end{equation}
where $k=1,2,\ldots$. These terms describe the prompt response rather than an additional family of QNM poles. The prompt expansion also includes a constant term.

We perturb this potential by adding a second P\"oschl-Teller bump,
\begin{align}
    V(x) &= \frac{V_A}{\cosh^2 \alpha_A x} + \frac{V_B}{\cosh^2 \alpha_B (x-x_B)}.
    \label{eq:PT_bump_potential}
\end{align}
Eq.~\eqref{eq:PT_bump_potential} describes a primary potential barrier $V_A$, modelling the photon-sphere barrier responsible for the original ringdown, while a weaker barrier $V_B$ is placed at $x_B>0$. 
We assume $V_B\ll V_A$ and $V_B/\alpha_B^2\ll1$. The ratio $V_B/\alpha_B^2$ controls the weak-barrier expansion, while the width of the perturbation is set by $\alpha_B^{-1}$. We also require $x_B\alpha_A\gg1$ and $x_B\alpha_B\gg1$, so that the overlap between the barriers can be neglected. Statements about a sharp first-echo time refer to this separated-barrier approximation. Exponentially small overlap contributions are neglected (this is the ``separated-barrier'' approximation). 
Here we have that $4 V_B < \alpha_B^2$. In this case, the expressions for $\gamma_n^B$, $\omega_n^B$ and $w_k^B$ are the same as for peak A (Eqs.~\eqref{eq:qnm_A},~\eqref{eq:lambda_A} and~\eqref{eq:defMatsubaraFreq}), but with $\lambda$ in the QNM frequency expression given by 
\begin{align}\label{eq:lambda_smallV}
\lambda_B &= \pm\frac{1}{2}   \sqrt{1-\frac{4 V_B}{\alpha_B^2}} ,
\end{align}
and the QNMs are purely decaying instead of oscillating and decaying. The $\pm$ sign determines the decay rate, and the QNMs contain a more quickly decaying and a more slowly decaying branch. In the limit where $V_B/\alpha_B^2 \rightarrow 0$, the fundamental (least-damped) QNM frequency of potential $B$ becomes
\begin{equation}
    \omega_0^B = -i\frac{V_B}{\alpha_B}\ .
\end{equation}
The least-damped nonconstant Matsubara term of potential $B$ has $i w_1^B=\alpha_B$. To leading order in $V_B/\alpha_B^2$, transmission through $B$ contains only the slowly decaying QNM. Reflection also contains a series of transient terms whose leading late-time decay scale is $\alpha_B^{-1}$, as shown in Appendix~\ref{App:list_greens_functions}.

\subsection{Propagation of a source through a potential}
\label{sec:propagation_source_prelim}

The Green's function encodes how a disturbance propagates through and scatters off a potential, allowing the response to arbitrary initial data to be constructed analytically~\cite{Leaver:1986gd,PhysRevD.55.468,Kuntz:2025gdq,Lagos_2023,Chavda:2024awq,Arnaudo:2025uos,DeAmicis:2025xuh,DeAmicis:2026wqd}. Consider initial data $\psi_0(x)=\psi_0(t_0,x)$ and $\dot{\psi}_0(x)=\partial_t\psi_0(t,x)|_{t=t_0}$. The retarded Green's function $G(x_1,x_0,t_1-t_0)$ propagates these data to the spacetime point $(t_1,x_1)$,
\begin{align} \label{eq:psi1}
    \psi_1(x_1, t_1) &= - \int_{- \infty}^\infty {\rm d} x_0 \big[ G(x_1,  x_0, t_1- t_0) \partial_{t_0}\psi_0(x_0,t_0) \nonumber \\
    &+ \partial_{t_1} G(x_1,  x_0, t_1-t_0) \psi_0( x_0,t_0) \big] \; .
\end{align}
Because $G$ is retarded, $G(x_1,x_0,t_1-t_0)=0$ whenever $t_1-t_0<|x_1-x_0|$. The integral in Eq.~\eqref{eq:psi1} then only has contributions from the intersection of the support of the initial data with the past light cone of ($t_1,x_1$). 
The Green's function itself is defined by the equation~\cite{Kuntz:2025gdq}
\begin{align}
    \left[-\partial_t^2 +\partial_x^2-V(x)\right]G(x,x'&,t-t') = \nonumber\\ & \delta(t-t')\delta(x-x'),
\end{align}
with $G=0$ for $t<t'$ and radiative boundary conditions at spatial infinity.
In the following we use the notation
\begin{align}
    &u = t-x, &v=t+x,
\end{align}
with subscripts to indicate different points $u_i=t_i-x_i$ (for example $i=0 \dots 3$ in Fig~\ref{fig:diagram_echo} in the next section). 

The Green's function for the transmission of initial data situated at a position $x_0$ on the left of a P\"oschl-Teller potential peak to an observer at $x_1$ to the right of the peak has the form
\begin{align}\label{eq:GA_full}
    G_A(x_1,x_0, &t_1- t_0) =\nonumber\\ & \Theta(u_1-u_0) G_A^{\rm QNM}(x_1,x_0, t_1-t_0).
\end{align}
where $G_A^{\rm QNM}$ contains only the QNMs of potential $A$ with no prompt response in the waveform, see~\cite{Kuntz:2025gdq,Arnaudo:2025uos}.

The large-separation assumption as described below Eq.~\eqref{eq:PT_bump_potential} ensures the existence of an intermediate region satisfying $x\alpha_A\gg1$ and $(x_B-x)\alpha_B\gg 1$. In this region, the exponentially decaying tails of both barriers are negligible. We can therefore organize the waveform as a multiple-scattering expansion, treating each interaction using the Green's function of the corresponding isolated P\"oschl-Teller potential. In this case, we can take $|x_0\alpha_A|\gg1$, $x_1\alpha_A\gg1$, and $(x_B-x_1)\alpha_B\gg1$. Then the Green's function from Eq.~\eqref{eq:GA_full} can be approximated to (see App.~\ref{App:list_greens_functions}):
\begin{align}
    G_A^{\rm QNM} (x_1, & x_0, t_1-t_0) \approx\text{Re} \sum_{n \geq 0} \gamma_{n}^A e^{i \omega_n^A (u_0-u_1)} ,
    \label{eq:GAqnm}
\end{align} 
where
\begin{align}
    \label{eq:gammaA}
    \gamma_n^A &= \frac{(-1)^{n+1}}{ n!} \frac{\Gamma \big( -2\lambda_A-n\big)}{\Gamma \big( \frac{1}{2} - \lambda_A - n \big)^2}.
\end{align}
In sums written with ${\rm Re}$, we use the branch with ${\rm Re}\,\omega_n^A>0$. Taking the real part includes its negative-frequency mirror.

Although our derivation formally assumes parametrically large separation, the asymptotic Green's function is already accurate at moderate distances. For the parameters used in Secs~\ref{sec:confirmation_numerics_timedomain} and ~\ref{sec:confirmation_numerics_freqdomain}, the modulus of the $n=1$ QNM radial factor in the exact Pöschl-Teller Green's function at $x\alpha_A=2$ differs from its asymptotic value by less than $1\%$.

Equation~\eqref{eq:GAqnm} is the transmission Green's function through barrier A needed below. The remaining configurations required in the multiple-scattering calculation (reflection from A, and transmission through and reflection from B) are listed in Appendix~\ref{App:list_greens_functions}. Green's functions with the observer on the opposite side of a barrier follow from the reflection symmetry $x-x_p\rightarrow-(x-x_p)$ about the corresponding peak $x_p$.

\section{Time domain solution using Green's function approach}\label{sec:time_domain_sol}

Here we calculate the time domain behaviour of a perturbation that starts as a delta function to the left of peak A ($x_0<x_A=0$), as seen by an observer to the right of peak B ($x_{\rm obs}>x_B)$. 

One component of the signal seen at $x_{\rm obs}$ has travelled through peak A, then directly through peak B and out to infinity. Other components have travelled through peak A and then reflected $n_{\rm echo}\geq 1$ times between peak B and peak A, before travelling through peak B and out to infinity. $n_{\rm echo}$ denotes the number of completed round trips in the cavity between A and B.

Here we calculate the parts of the signal seen by an observer, $\psi_{\rm obs}$, with $n_{\rm echo}=0, 1$. This calculation already provides insight into the connection between the time domain and the frequency domain behavior. We show here only the terms that contribute to the change in signal frequencies. 

The paths for $n_{\rm echo}=0,1$, and the points at which the behaviour is calculated are described in Fig.~\ref{fig:diagram_echo}.

\subsection{Component with $n_{\rm echo}=0$ }
\label{sec:n=0}

Here we consider the component that transmits through peak $A$ and then through peak $B$.

We start with an initial perturbation to the left of peak $A$ given by
\begin{align}
    &\psi_0(x) =0 & \dot \psi_0(x) = -\delta(x-x_0).
\end{align}
After transmitting through peak A, the field is simply given by the QNM part of the Green's function,
\begin{equation}\label{eq:field_through_A}
    \psi_1(t_1, x_1) = \Theta(u_1-u_0) {\rm Re} \sum_n \gamma_n^A e^{i \omega_n^A (u_0 - u_1)}.
\end{equation}
This means that transmitting through peak A results in a pure QNM ringing, with the QNM frequencies of peak A.

We use Eq.~\eqref{eq:psi1} with $\psi_1$ instead of $\psi_0$ as the initial condition to understand the subsequent transmission through peak B. This gives 
\begin{align}
\label{eq:psiobs_n0}
    &\psi(x_{\rm obs}, t_{\rm obs}) =   {\rm Re} \sum_{n}  \gamma_n^A   \bigg( (1 - i \epsilon_n)  e^{i \omega_n^A(u_0-u_{\rm obs})} \nonumber \\
    &+ i \epsilon_n   e^{\frac{V_B}{\alpha_B} (u_0-u_{\rm obs})}  \bigg)   \Theta(u_{\rm obs}-u_0) +\mathcal{O}(\epsilon_n^2).
\end{align}
with $\epsilon_n = V_B/(\alpha_B \omega_n^A)$ and $|\epsilon_n|\ll1$. This second transmission does not shift the existing QNM frequencies of A. It modifies their amplitudes at $O\left(\epsilon_n\right)$, and introduces the purely decaying QNMs of peak B, $i\omega^B_0 = V_B/\alpha_B$, with amplitudes of order $O\left(\epsilon_n\right)$. Notice that the QNM spectrum of potential B contains many frequencies, but only the slow decaying mode proportional to $V_B$ has a nonzero amplitude to $\mathcal{O}(\epsilon_n)$, with the rest of the spectrum sitting in the $\mathcal{O}(\epsilon_n^2)$ term.

Direct transmission through $B$ preserves the frequency of each $A$-QNM component, while changing its amplitude and adding contributions associated with $B$. We now examine the first return to $A$, where a second interaction with peak A produces a resonant contribution.


\subsection{Component with $n_{\rm echo}=1$ }\label{sec:n=1}

We now consider the component of the waveform that transmits through A, scatters once from B, scatters from A, and then transmits through B. We use the points $0-3$ and an observer point as indicated in Fig.~\ref{fig:diagram_echo}, where $x_0<x_A$, $x_A<x_1<x_B$, $x_A<x_2<x_B$, $x_A<x_3<x_B$, and $x_{\rm obs}>x_B$. 
As in Sec.~\ref{sec:n=0}, the field $\psi_1$ after transmitting through A is simply given by equation~\eqref{eq:field_through_A}

We next compute the field $\psi_2$, after scattering from $B$. As $\psi_1$ is travelling to the right ($\partial_{t_1} \psi_1 = - \partial_{x_1} \psi_1$), we can integrate by parts the Green's function formula to have
\begin{equation}
    \psi_2 = - \int {\rm d}x_1 \big[ \partial_{x_1} G_B + \partial_{t_2} G_B \big] \psi_1 \; .
\end{equation}
Doing the explicit computation (see Appendix~\ref{App:n=1calc}), we find that there is a transient (prompt) part of $\psi_2$ in which we are not interested, while the waveform after $v_2>u_0+2x_B$ is given by
\begin{align}
    &\psi_2 = \Theta(v_2-u_0-2x_B)   {\rm Re}\sum_n i\gamma_n^A  \epsilon_n \Bigg[     e^{ \frac{V_B}{\alpha_B}(u_0-v_2+2x_B)}\nonumber \\
    &- {}_2F_1 \bigg(1, i \frac{\omega_n^A}{\alpha_B} , 1 + i \frac{\omega_n^A}{\alpha_B}, - e^{\alpha_B(v_2-u_0-2x_B)} \bigg)  \Bigg] + \mathcal{O}(\epsilon_n^2) \,,
\end{align}
where ${}_2F_1$ is a hypergeometric function.
As before, $\epsilon_n = V_B/(\alpha_B \omega_n^A)$ and $|\epsilon_n|\ll1$. The reflected waveform is of small amplitude, as expected. The meaning of the time $v_2>u_0+2x_B$ after which the QNMs of potential $B$ start showing up in the waveform, is simply related to the fact that the wave has to \textit{bounce} off $x=x_B$ before picking up the QNM frequencies of $V_B$.

To understand the frequency content of $\psi_2$, we notice that the hypergeometric function at late times (for $\alpha_B(v_2-u_0-2x_B)\gg1$, i.e. some time after the bounce time) can be approximated in closed form. Its leading transient decays as $e^{-\alpha_B(v_2-u_0-2x_B)}$, while another term oscillates at the QNM frequencies of potential $A$. This term has the form 
\begin{align}
{}_2F_1&\left(
1,
\frac{i\omega_n^A}{\alpha_B},
1+\frac{i\omega_n^A}{\alpha_B},
-e^{\alpha_B(v_2-u_0-2x_B)}
\right)\nonumber\\ &
\supset 
\frac{\pi i\omega_n^A/\alpha_B}
{\sin\left(\pi i\omega_n^A/\alpha_B\right)}
e^{-i\omega_n^A(v_2-u_0-2x_B)}. 
\end{align}
In summary,  the part of $\psi_2$ which oscillates at the QNM frequency of potential $A$ is, for large enough times,
\begin{align}
    \psi_2 &\supset
\Theta(v_2-u_0-2x_B)
{\rm Re}\sum_n
\gamma_n^A \epsilon_n
e^{i\omega_n^A(u_0-v_2+2x_B)}\nonumber\\ &\times
\left[
\frac{\pi \omega_n^A/\alpha_B}
{\sin\left(\pi i\omega_n^A/\alpha_B\right)}
\right] + \mathcal{O}(\epsilon_n^2) .  \label{eq:psi2QNMA}
\end{align}
We will only need this part of $\psi_2$ to understand the origin of the spectral instability. 
We are now ready for the last reflection on potential $A$, giving us the field $\psi_3$. This is the interaction that produces the resonance which effectively shifts the observed frequency. Since now $\psi_2$ is a wave travelling to the left, we have the formula
\begin{equation} \label{eq:psi3Green}
    \psi_3 =  \int {\rm d}x_2 \big[ \partial_{x_2} G_A - \partial_{t_3} G_A \big] \psi_2 \; .
\end{equation}
Nonresonant contributions contain the characteristic frequencies of the individual barriers. The contribution involving the same $A$-QNM frequency in both the incident field and the Green's function must be treated separately. When $n=m$, the integrand in Eq.~\eqref{eq:psi3Green} is independent of $x_2$. The integral then produces a term with a secular, linear-in-time prefactor proportional to the length of the integration domain.
 To find this term we need to convolve the QNM Green's function of potential A with the part of the field $\psi_2$ written in Eq.~\eqref{eq:psi2QNMA}. An explicit computation in Appendix~\ref{App:n=1calc} gives 
\begin{align} \label{eq:psi3Resonant}
    \psi_3 &\supset  (u_3 - u_0 - 2x_B) \Theta(u_3 - u_0 - 2x_B) {\rm Re}\sum_n (-1)^n \omega_n^A\nonumber \\ & \times (\gamma_n^A)^2 \epsilon_n  e^{i \omega_n^A(u_0-u_3+2x_B)} \bigg( \frac{\pi i\omega_n^A/\alpha_B}
{\sin\left(\pi i\omega_n^A/\alpha_B\right)} \bigg) + \mathcal{O}(\epsilon_n^2) .  
\end{align}
It is clear from the above that only this resonant combination of terms has this secular prefactor $\propto t_3$. Note that the exponential QNM decay will eventually win over the linear prefactor to drive the waveform to zero at late times. This is the secular contribution that we use to relate the first echo to the perturbative pole shift. 

Finally, as we saw in Sec.~\ref{sec:n=0}, the transmission through peak B changes the amplitude of this component and adds the characteristic modes of peak B. These are relative $\mathcal{O}(\epsilon_n)$ modifications to a waveform which is already a small $\mathcal{O}(\epsilon_n)$ quantity, making those changes $\mathcal{O}(\epsilon_n^2)$. Therefore, to $\mathcal{O}(\epsilon_n)$, $\psi_{\rm obs}$ contains exactly the resonant term written in Eq.~\eqref{eq:psi3Resonant}.

\subsection{Combination of terms}

The linear-in-time prefactor in the $n_{\rm echo}=1$ component describes how the spectral instability appears in the waveform. In the small-shift regime, it can be interpreted as a perturbation to the frequencies of potential $A$. To see this, we combine the terms oscillating at the QNM frequencies of potential $A$ in the $n_{\rm echo}=1$ term with the $n_{\rm echo}=0$ term to find 
\begin{widetext}    
\begin{align}
    \label{eq:psiobs}
    \psi_{\rm obs} &\supset \Theta(u_{\rm obs}-u_0){\rm Re} \sum_n  \gamma_n^A (1 - i \epsilon_n)   e^{i \omega_n^A  (u_0 - u_{\rm obs} )}\nonumber \\ &+ (u_{\rm obs} - u_0 - 2x_B) \Theta(u_{\rm obs} - u_0 - 2x_B)   {\rm Re}\sum_n (-1)^n \omega_n^A   (\gamma_n^A)^2 \epsilon_n e^{i \omega_n^A(u_0-u_{\rm obs}+2x_B)} \bigg( \frac{\pi i\omega_n^A/\alpha_B}
{\sin\left(\pi i\omega_n^A/\alpha_B\right)} \bigg) + \mathcal{O}(\epsilon_n^2)\ .
\end{align}
\end{widetext}
Equation~\eqref{eq:psiobs} retains the prompt $A$-QNM contribution and the resonant part of the first echo. Some nonsecular first-order contributions are omitted.

After the first echo, $u_{\rm obs}>u_0+2x_B$, the secular response can be approximated by a shifted exponential when $|\delta\omega_n^A(u_{\rm obs}-u_0-2x_B)|\ll1$. We write 
\begin{align}
    \psi_{\rm obs} \supset {\rm Re} \sum_n & \gamma_n^A (1 - i \epsilon_n)     e^{i \omega_n^A  (u_0 - u_{\rm obs} ) + i \delta \omega_n^A(u_0 - u_{\rm obs} + 2x_B)} \nonumber\\ \simeq {\rm Re} \sum_n &\gamma_n^A  (1 - i \epsilon_n)    e^{i \omega_n^A  (u_0 - u_{\rm obs})} \nonumber\\ & \times (1 + i \delta \omega_n^A  (u_0-u_{\rm obs} + 2 x_B))\ .
\end{align}
To match Eq.~\eqref{eq:psiobs}, in the small frequency shift limit, we have 
\begin{equation}
     \label{eq:delta_omega}
    \delta \omega_n^A =  (-1)^{n+1} \omega_n^A \gamma_n^A \epsilon_n e^{2 i \omega_n^A x_B}  \bigg(
\frac{\pi \omega_n^A/\alpha_B}
{\sin\left(\pi i\omega_n^A/\alpha_B\right)}\bigg) \, .
\end{equation}

This formula matches results from the literature, in that it predicts that the frequency shift goes to zero as $\epsilon_n \rightarrow0$, increases exponentially with $x_B$, and affects different overtones differently~\cite{Cardoso:2024mrw,Ianniccari:2024ysv,Yang:2024vor,Berti:2022xfj}. It is worth stressing that the magnitude of the deviation is weighted by the exponential, which makes higher modes acquire a much larger modification than the fundamental mode.

The secular contribution turns on after one round trip, at $u_{\rm obs}-u_0=2x_B$. In the small-shift regime, its coefficient reproduces the leading frequency-domain pole shift, as shown below. Outside that regime, it describes a finite-time waveform correction rather than an instantaneous replacement of the original QNM frequency by a perturbed one.

The shifted-exponential interpretation uses $|\delta\omega_n^A(u_{\rm obs}-u_0-2x_B)|\ll1$, but the terms retained in Eq.~\eqref{eq:psiobs} can describe the fitted frequency evolution beyond that limit. We test this during the first-echo window, $2x_B<u_{\rm obs}-u_0<4x_B$, before the second echo arrives. As shown in Sec.~\ref{sec:confirmation_numerics_freqdomain}, fitting these terms gives effective frequencies close to those obtained from the numerical waveform.

\section{Confirmation of results}
\subsection{Comparison to literature}\label{sec:confirmation_literature}

We now compare the result of Eq.~\eqref{eq:delta_omega} to the analytic frequency domain results of Refs.~\cite{Ianniccari:2024ysv,Yang:2024vor}. Ref.~\cite{Ianniccari:2024ysv} considers a wave equation with a primary Schwarzschild Regge-Wheeler peak, as well as a secondary, perturbing, P\"oschl-Teller peak. While exact QNMs can be solved for an arbitrary potential numerically, Ref.~\cite{Ianniccari:2024ysv} gives a closed form, analytic expression for the shift in frequencies $\delta\omega$ due to the presence of the secondary bump. While the potential considered in that reference is not the double P\"oschl-Teller one considered in this work, their transfer-matrix expression for the migration of a QNM pole is general for two sufficiently separated localized potentials, and can therefore be compared to our system. 

In order to write the equation for the shift in QNMs due to the presence of a bump, Ref.~\cite{Ianniccari:2024ysv} assumes a weak bump, $V_B/\alpha_B^2\ll1$, negligible overlap between the two potential barriers, and a small frequency shift compared with the unperturbed frequency, $|\Delta\omega_n/\omega_n^A|\ll1$. Ref.~\cite{Ianniccari:2024ysv} assumes the hierarchy $V_B/\alpha_B^2\ll|\Delta\omega_n/\omega_n^A|\ll1$, so the frequency migration can be larger than the perturbation, while remaining within the small-shift regime. Because the shift grows exponentially with $x_B$, this approximation eventually breaks down at sufficiently large separations. These conditions inform the frequency-domain interpretation of Eq.~\eqref{eq:delta_omega}.

We find that the result in Eq.~\eqref{eq:delta_omega} is  algebraically equivalent to the approximated closed form result for $\delta\omega$ in Ref.~\cite{Ianniccari:2024ysv} after translating the notation. In the notation of Ref.~\cite{Ianniccari:2024ysv}, the leading
frequency shift of a pole $\omega_n$ is
\begin{equation}\label{eq:Ianniccari_shift}
    \Delta\omega_n =
    e^{2 i \omega_n c}
    \left.\operatorname{Res} r_A'\right|_{\omega_n}\epsilon \kappa_B\left(\omega_n\right).
\end{equation}
In our notation, 
\begin{equation}
    \epsilon=\frac{V_B}{\alpha_B^2},\qquad
    c=x_B,\qquad
    \kappa_B(\omega)=-\frac{i\pi}
    {\sinh(\pi\omega/\alpha_B)}.
\end{equation}
In Eq.~\eqref{eq:Ianniccari_shift} the reflection coefficient of the weak perturbing P\"oschl-Teller potential is $r_B = \epsilon\kappa_B(\omega)+\mathcal{O}(\epsilon^2)$, and $r_A'$ is the reflection coefficient of the primary potential, where the prime on $r_A'$ indicates the reflection of the incident wave is from the right. Near the $n$th QNM pole of the primary potential,
\begin{equation}
r_A'(\omega)
=
\frac{
\operatorname{Res}_{\omega=\omega_n^A}r_A'
}{\omega-\omega_n^A}
+\mathcal{O}(1).
\end{equation}
And so, from Eq.~\eqref{eq:Ianniccari_shift}, the interaction of a QNM of potential $A$ with peak $B$ enters through its reflection amplitude, and the response of $A$ is determined by the residue of its QNM pole. 

Next we evaluate $\left.\operatorname{Res} r_A'\right|_{\omega_n}$, the residue of the reflection coefficient at the $n$th QNM pole. Using the reflection coefficient given in the Supplemental Material of Ref.~\cite{Ianniccari:2024ysv}, with $r_s=\alpha_A^{-1}$ and $\lambda_I=1/2+\lambda_A$, where $\lambda_I$ denotes the parameter $\lambda$ used in Ref.~\cite{Ianniccari:2024ysv}, the QNM pole at $\omega=\omega_n^A$ arises from the factor
\begin{equation}
    \Gamma(1/2+\lambda_A-i\omega/\alpha_A).
\end{equation}
Since $1/2+\lambda_A-i\omega/\alpha_A=-n$, and using $\operatorname{Res}_{z=-n}\Gamma(z) = (-1)^n/n!$, together with the Gamma-function  recurrence and reflection identities, we obtain
\begin{equation}
    \left.\operatorname{Res}r_A'\right|_{\omega_n^A}=(-1)^{n+1}\omega_n^A\gamma_n^A .
\end{equation}
Substituting these expressions into Eq.~\eqref{eq:Ianniccari_shift} gives
\begin{equation}
    \Delta\omega_n =
    i(-1)^n\frac{V_B}{\alpha_B^2}
    \frac{\pi\omega_n^A\gamma_n^A}
    {\sinh(\pi\omega_n^A/\alpha_B)}
    e^{2i\omega_n^A x_B},
\end{equation}
which is identical to Eq.~\eqref{eq:delta_omega}.

Ref.~\cite{Ianniccari:2024ysv}  compares their perturbative result with the numerical QNM spectrum obtained in Ref.~\cite{Cheung:2021bol} for a Schwarzschild potential perturbed by a P\"oschl--Teller bump. For the representative choice $V_B/\alpha_B^2=10^{-6}$, the analytic closed form approximation reproduces the characteristic spiralling shifts of the fundamental mode in the complex-frequency plane, with errors at the $\sim10\%$ level. The perturbative expression applies while the migrated mode remains close to the corresponding unperturbed pole, $\epsilon\ll|\Delta\omega_n/\omega_n|\ll1$. At sufficiently large separations, the exponential frequency shift violates this condition, and the system enters a nonperturbative regime. 

The connection between the frequency- and time-domain calculations can be understood from the multiple-scattering expansion. At first order in the reflection from potential $B$, the product of the QNM poles of the transmission and reflection coefficients of potential $A$ produces a double pole $(\omega-\omega_n^A)^{-2}$. In the time domain, this double pole gives the secular contribution $(t-t_{\rm echo})e^{-i\omega_n^A(t-t_{\rm echo})}$ identified in Eq.~\eqref{eq:psiobs}, with $t_{\rm echo}=2x_B$. The full multiple-scattering series determines the poles of the combined system, whereas a finite-time waveform contains only the contributions that have had time to arrive. Matching the first secular coefficient identifies the leading perturbative pole shift without requiring a late-time limit of the waveform.

The resonant first-echo contribution turns on after one round trip in the separated-barrier approximation. Its interpretation as a frequency-domain pole shift requires the small-shift conditions stated above. Similar causal delays of the perturbed-QNM contribution were identified in Refs.~\cite{Ianniccari:2024ysv,Yang:2024vor}. Our result determines the behaviour before and after the first echo for all overtones. We have shown that the resonant time-domain secular term is related precisely to the frequency-domain pole migration.

\subsection{Confirmation of causal behavior of Eq.~\eqref{eq:psiobs} with numerical evolution}
\label{sec:confirmation_numerics_timedomain}
Eq.~\eqref{eq:psiobs} predicts that the resonant, secular $n_{\rm echo}=1$ contribution appears one round-trip time, $\Delta t= 2(x_B-x_A)=2x_B$, after the prompt ringdown. If this term is responsible for the frequency shifts, a time-domain frequency fit should recover the unperturbed spectrum before this time and begin to depart from it only after the first echo.

A related time-domain numerical analysis was performed in Ref.~\cite{Berti:2022xfj}, where the prompt fundamental mode was found to remain close to its unperturbed value, while the perturbed spectrum emerged only after the echo-delayed response. That analysis focused on the fundamental mode and left the corresponding behavior of the overtones unresolved. Here we perform a similar test using initial data designed to enhance the $n=1$ overtone.

We numerically evolve Eq.~\eqref{eq:RWZeq} with $V_A = \alpha_A^2$, $x_A = 0$ and varying bump $B$ parameters. As initial data we use the solutions in~\cite{Berti:2009kk} to construct an overtone dominated solution and then apply a window function to suppress the growth as $x\rightarrow\infty$. We choose initial data dominated by the $n=1$ overtone for two reasons. First, for the parameter choices we explore, the overtone shift is much larger than that of the fundamental mode, making the causal transition easier to resolve. Second, we want to show that this causal behaviour is valid in the case of overtones as well as the fundamental mode. The particulars of the initial data and evolution scheme are detailed in Appendix~\ref{App:numerics}.

We measure the perturbation as seen by a distant observer at $x_{\rm obs}\alpha_A=80$, and fit the output using \href{https://mhycheung.github.io/jaxqualin/}{Jaxqualin}~\cite{Cheung:2023vki}. 
We fit using a model  
\begin{align}
    \label{eq:free_freq_model}
    \psi = \sum_m A_m e^{-i\omega_m (t-t_{\rm start})}\; ,
\end{align}
where $A_m \in \Bbb{C}$ and $\omega_m\in \Bbb{C}$ are free parameters. $t_{\rm start}$ is the fit start time. So in total there are $4m$ free parameters in the model where $m$ is the number of modes. 
For the parameter choice considered below, the early-time fit clusters around the unperturbed $n=1$ frequency, whereas after the first-echo time it migrates toward the perturbed frequency-domain prediction.

\begin{figure}[h]
\centering
\includegraphics[width=9cm]{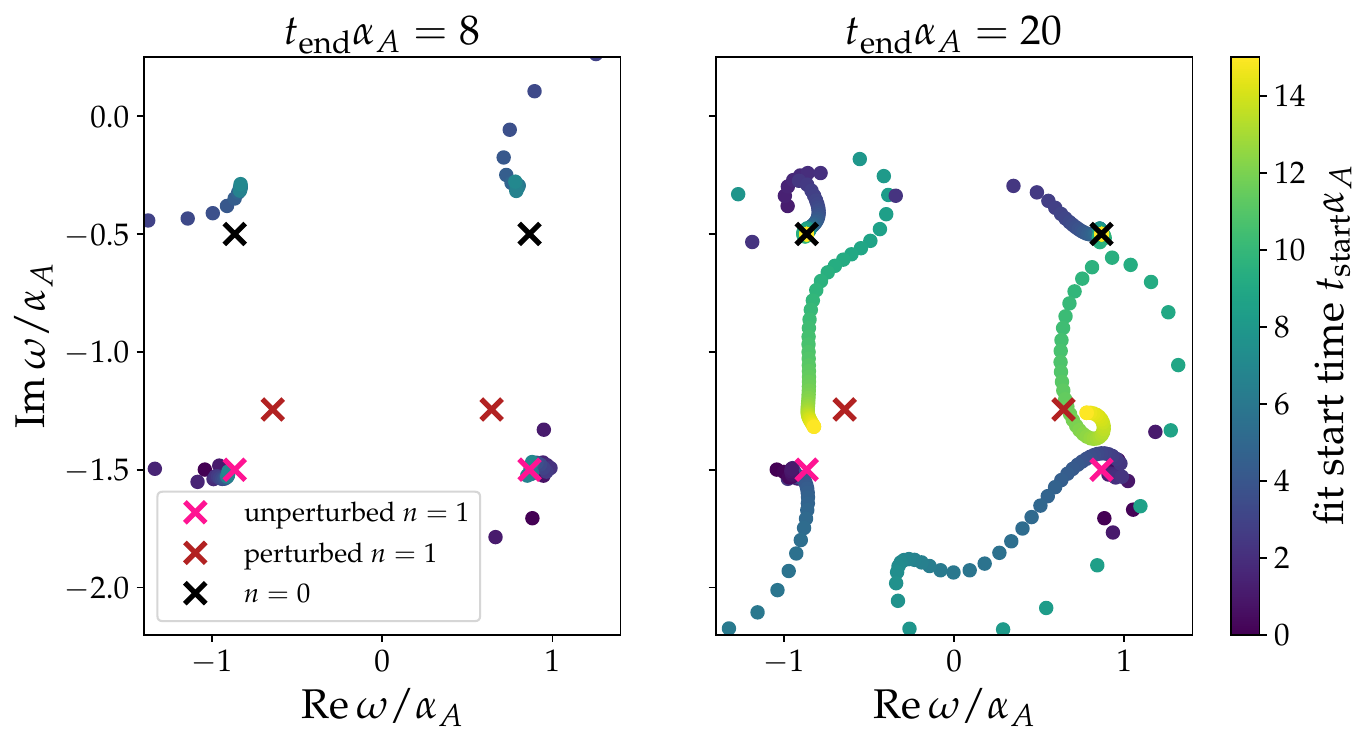}
\caption{Free-frequency fits to the numerical waveform as a function of the fit start time $t_{\rm start}$, using four free modes and fixed end times $t_{\rm end}\alpha_A=8$ (left) and $t_{\rm end}\alpha_A=20$ (right). The potential parameters are $V_A=\alpha_A^2=1$, $x_B\alpha_A=4$, $V_B/\alpha_A^2=10^{-4}$, and $\alpha_B/\alpha_A=2$. Both members of each positive- and negative-frequency mirror pair are included in the fit. Crosses indicate the unperturbed and perturbed frequency-domain QNM predictions, computed with the method outlined in Appendix~\ref{App:freq}. For this example, the first-echo time is $t\alpha_A=2x_B\alpha_A=8$. 
}
\label{fig:free_freq_fit}
\end{figure}

In Fig.~\ref{fig:free_freq_fit} we show the results of a free frequency fit on data where $V_A=\alpha_A^2$, $V_B = 0.0001\alpha_A^2$, $\alpha_B=2\alpha_A$, $x_B=4/\alpha_A$. We shift the numerical time coordinate such that $t_{\rm ring}=0$, where $t_{\rm ring}$ denotes the onset of the clean ringdown signal at the observer, as defined in Appendix~\ref{App:initial_data}. All times quoted below are measured relative to this reference time. We report times and lengths in units of $\alpha_A^{-1}$, frequencies in units of $\alpha_A$, and potential amplitudes in units of $\alpha_A^2$.

We perform fits over windows of data at times $t\in [t_{\rm start},t_{\rm end}]$.  We perform the fit with two different fixed end times. Choosing $t_{\rm end}\alpha_A=2x_B\alpha_A=8$, corresponding to the first-echo time, restricts the fitting interval to the pre-echo waveform and therefore tests whether the original $A$-QNM components retain their frequencies in the early ringdown. We then extend the fitting interval to $t_{\rm end}\alpha_A=5x_B\alpha_A=20$, so that it contains the post-echo signal. In this case, fits beginning before the echo generally contain both the pre- and post-echo portions of the waveform, and the recovered frequencies should be interpreted as effective frequencies over the entire fitting interval.

The fit is shown with varying start time, to demonstrate the stability of the frequency fit to small changes in the signal. Isolated fitted frequencies that vary strongly under small changes in the fit start time are not interpreted as robust mode detections. Clusters that remain approximately stationary over a range of neighboring start times identify stable fitted components. 

From Fig.~\ref{fig:free_freq_fit}, the trajectory of the effective frequency fit for $t_{\rm end}\alpha_A=2x_B\alpha_A=8$ never deviates from the unperturbed overtone. Because the initial data are constructed to enhance the $n=1$ overtone, the fundamental mode is small over the $t_{\rm end}\alpha_A=8$ fitting interval and its recovered frequency is less stable. 

For $t_{\rm end}\alpha_A=20$, the fitted overtone begins to depart from the unperturbed value at $t_{\rm start}\alpha_A\approx6$. Since these fitting windows extend beyond the first-echo time $t\alpha_A=8$, they contain both pre- and post-echo portions of the waveform. The corresponding intermediate frequencies should be interpreted as effective finite-window fits rather than as evidence for a frequency shift before the echo arrives.

The numerical evolution therefore confirms the causal structure predicted by Eq.~\eqref{eq:psiobs}. A fitting window restricted to times up to the first echo recovers the unperturbed overtone, showing that the fitted overtone retains its unperturbed frequency before the echo. Once the fitting interval contains the first reflected signal, the recovered effective frequency begins to migrate, and a fit restricted to post-echo times approaches the perturbed-spectrum prediction. This supports the identification of the resonant $n_{\rm echo}=1$ contribution as the mechanism responsible for the observed spectral migration.

\subsection{Comparing effective frequencies from Eq.~\eqref{eq:psiobs} with numerical evolution}\label{sec:confirmation_numerics_freqdomain}

The small-frequency-shift approximation used to obtain Eq.~\eqref{eq:delta_omega} is not required in deriving the first-echo waveform in Eq.~\eqref{eq:psiobs}. We therefore ask whether Eq.~\eqref{eq:psiobs} continues to describe the finite-time frequency content of the numerical waveform when the secular correction can no longer be interpreted as a small shift of the unperturbed QNM. We test this in a parameter regime for which the bump remains weak $V_B/\alpha_B^2\ll1$, but $\left|\delta\omega^A_1/\omega^A_1\right| \sim 1 $, so that Eq.~\eqref{eq:delta_omega} is outside its regime of validity.

We fit the expression from Eq.~\eqref{eq:psiobs}, evaluated over a grid in time, as well as the output of a numerical evolution, using \href{https://mhycheung.github.io/jaxqualin/}{Jaxqualin}~\cite{Cheung:2023vki}. For the analytic comparison we isolate the complex $n=1$ component of Eq.~\eqref{eq:psiobs} before taking the real part, and therefore fit a single free complex frequency. For the real numerical waveform, both members of the mirror pair $\omega$ and $-\bar{\omega}$ are present (as can be seen in Fig.~\ref{fig:free_freq_fit}). We impose this relation between the frequencies while leaving the two complex amplitudes independent. The resulting complex-valued model is fitted to the real data. Explicitly, we use
\begin{align}
    \label{eq:free_freq_model_symm}
    \psi = \sum_n \left(A_n e^{-i\omega_n (t-t_{\rm start})}+B_n e^{+i\bar{\omega}_n (t-t_{\rm start})}\right)\; ,
\end{align}
where the free parameters are $A_n\in \Bbb{C}$, $B_n\in \Bbb{C}$  and $\omega_n\in\Bbb{C}$.
This better conditions the fit and leads to more precise results. We use a single free frequency for the analytical case, with no mirror symmetry imposed, using the model from Eq.~\eqref{eq:free_freq_model}. We use the model from Eq.~\eqref{eq:free_freq_model_symm} with two pairs of frequencies for the numerical evolution fit (one pair for the $n=0$ and its mirror, and one for the $n=1$ and its mirror). We apply this more constrained fitting procedure to the same numerical evolution shown in Fig.~\ref{fig:free_freq_fit}.

\begin{figure}[h]
\centering
\includegraphics[width=9cm]{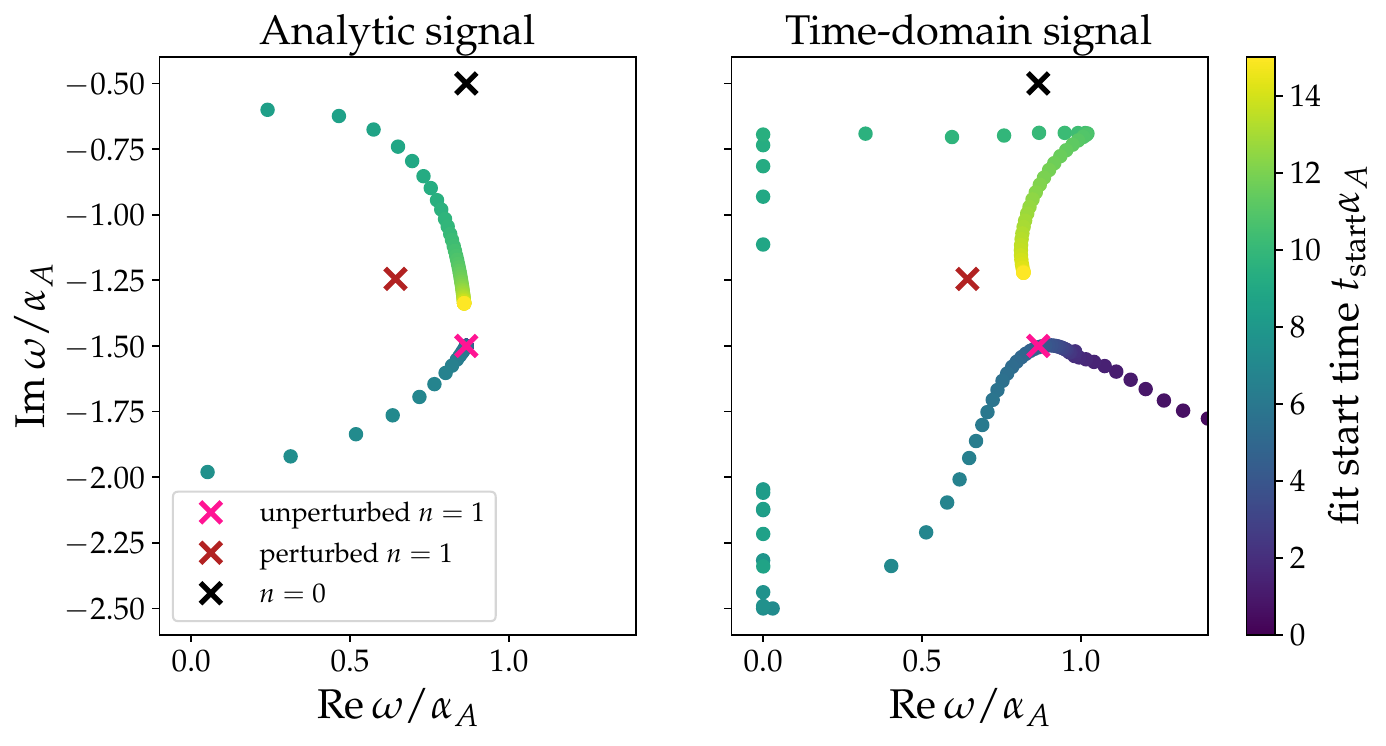}
\caption{Comparison of the effective $n=1$ overtone frequency obtained from the analytic first-echo waveform, Eq.~\eqref{eq:psiobs} (left), and from the full numerical evolution (right), as a function of the fit start time $t_{\rm start}$. In the analytic case, we fit the isolated $n=1$ component with a single free complex frequency. In the numerical case, two pairs of mirror frequencies are fitted with independent complex amplitudes. Only the pair associated with the $n=1$ overtone is shown, while the second pair remains close to the fundamental $n=0$ mode and is not included in the plot for clarity. The potential parameters are $V_A=\alpha_A^2$, $V_B=10^{-4}\alpha_A^2$, $\alpha_B=2\alpha_A$, and $x_B=4/\alpha_A$, as in Fig.~\ref{fig:free_freq_fit}. Crosses indicate the unperturbed and perturbed frequency-domain QNM predictions. The first echo arrives at $t\alpha_A=2x_B\alpha_A=8$. Here $t_{\rm end}\alpha_A=4x_B\alpha_A=16$, so the fit is performed before the second echo contributes, while Eq.~\eqref{eq:psiobs} is still valid.
}
\label{fig:free_freq_fit_comp}
\end{figure}

In Fig.~\ref{fig:free_freq_fit_comp} we show the results of this analysis, where the end of the fitting window is fixed to $t_{\rm end}\alpha_A=4x_B\alpha_A=16$. As in Fig.~\ref{fig:free_freq_fit}, $V_A = \alpha_A^2$, $V_B = 0.0001\alpha_A^2$, $\alpha_B=2\alpha_A$, $x_B=4/\alpha_A$. For this case, Eq.~\eqref{eq:delta_omega} predicts that $\left|\delta\omega^A_1/\omega^A_1\right| \simeq 0.996$, clearly violating the approximation $\left|\delta\omega/\omega\right| \ll 1$ for the first overtone. The weak-bump approximation remains well satisfied, $V_B/\alpha_B^2 = 2.5\times10^{-5}\ll1$. The barriers are also spatially separated, although $x_B\alpha_A=4$ is only moderately large.

As in Sec.~\ref{sec:confirmation_numerics_timedomain}, we shift the numerical time coordinate such that $t_{\rm ring}=0$, where $t_{\rm ring}$ denotes the onset of the clean ringdown signal at the observer, as defined in Appendix~\ref{App:initial_data}. The effective frequencies recovered from the numerical waveform closely resemble those obtained by fitting the $n=1$ component of Eq.~\eqref{eq:psiobs} throughout the first-echo interval.

While the results are close, they are not identical. The initial data in these two cases is not the same, and so the fits are not expected to be identical either. In the analytic case the initial data is a delta function, where we have taken only the $n=1$ first overtone component after the first interaction with peak A. In addition, the analytic terms used for this comparison are only those given in Eq.~\eqref{eq:psiobs}. In the numerical evolution case, the initial data is set to be as close as possible to a pure $n=1$ overtone solution to the unperturbed potential, as described in Appendix~\ref{App:initial_data}. For the initial data and parameter choices studied here, we find that the relative fundamental-mode amplitude is approximately proportional to $V_B/V_A$. Its slower decay eventually limits the interval over which the overtone can be fitted reliably. 

The agreement shows that the resonant secular term captures the observed spectral migration during the first-echo window even when it cannot be interpreted as a time-independent shift of the QNM frequency. In this regime, the fitted frequencies should be understood as effective finite-time frequencies of the waveform rather than as the poles of the perturbed system.

\section{Characterizing the spectral instability using time-domain data}

In Sec.~\ref{sec:confirmation_numerics_freqdomain} we characterized the spectral instability using the free-frequency models of Eqs.~\eqref{eq:free_freq_model} and~\eqref{eq:free_freq_model_symm}. Related secular fitting models have been used to describe QNM resonances near exceptional points~\cite{Yang:2025dbn,Imafuku:2026rpn}. Here the secular term instead has a causal onset fixed by the echo delay. We therefore consider an alternative model motivated by the secular contribution in Eq.~\eqref{eq:psiobs},
\begin{align}
\label{eq:free_freq_model_sec}
\psi = \sum_m \Big[
&A_m+B_m(t-t_{\rm ring}-2x_B)
\Theta(t-t_{\rm ring}-2x_B)
\Big]\nonumber\\
&\times e^{-i\omega_m(t-t_{\rm ring})},
\end{align}
where $A_m,B_m\in\mathbb{C}$, $x_B\in\mathbb{R}$, and $\omega_m\in\mathbb{C}$. Here $t_{\rm ring}$ is a fixed reference for the onset of the ringdown, so that $2x_B$ measures the delay between the ringdown and the onset of the secular response.

We fit the same numerical data from Sec.s~\ref{sec:confirmation_numerics_timedomain} and~\ref{sec:confirmation_numerics_freqdomain} using this model.
Fitting all of the parameters in Eq.~\eqref{eq:free_freq_model_sec} simultaneously is poorly conditioned. In particular, $B_1$ and $\omega_1$ are strongly degenerate, since a small change in frequency generates a term linear in time at leading order. We fix the frequencies to their unperturbed values, and impose the relation $\omega\leftrightarrow-\bar{\omega}$ between the frequencies as before, without constraining the amplitudes of each pair. For the example considered here, we set $B_0=0$. The fundamental mode is then included in the fit without a secular correction, while a secular term is associated with the $n=1$ overtone. The same procedure could in principle be applied successively to other overtones.

\begin{figure}[h]
\centering
\includegraphics[width=7cm]{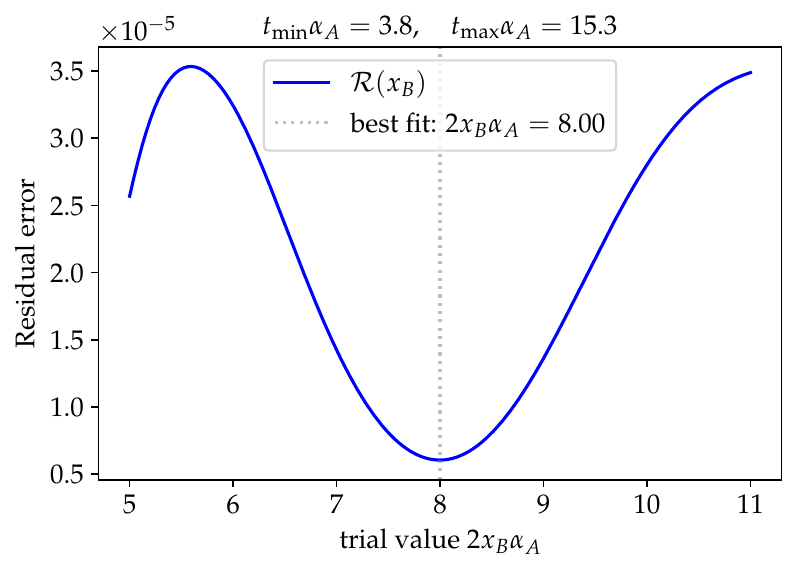}
\caption{Normalized fit residual as a function of the trial echo delay $2x_B$. The minimum determines the preferred value of $2x_B$ for this fitting window.}
\label{fig:residual_xB}
\end{figure}

Rather than fitting $x_B$ directly, we profile over it. For each trial value of $x_B$, the remaining complex amplitudes are determined by linear least squares, and we evaluate
\begin{equation}
\mathcal{R}(x_B)= \sum_{t=t_{\rm min}}^{t_{\rm max}}
\frac{|\psi-\psi_{\rm fit}|^2}{|\psi|^2}.
\end{equation}
The minimum of $\mathcal{R}(x_B)$ defines the preferred echo delay $2x_B$. An example of this profile is shown in Fig.~\ref{fig:residual_xB}.

\begin{figure}[h]
\centering
\includegraphics[width=9cm]{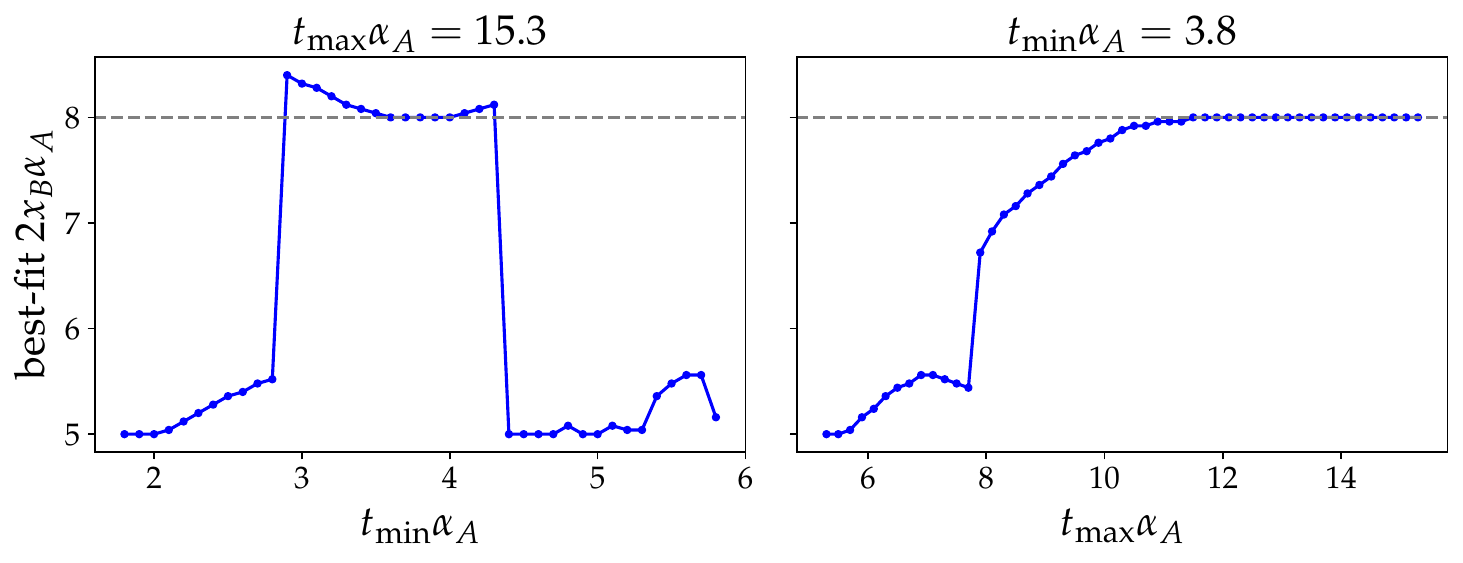}
\caption{Stability of the inferred echo delay $2x_B$ under variations of the fitting window. Left: varying $t_{\min}$ at fixed $t_{\max}$ reveals a range over which the fitted value is stable. Right: varying $t_{\max}$ at fixed $t_{\min}$ shows that the result converges once sufficiently much post-echo data are included.}
\label{fig:xB_with_window}
\end{figure}

The inferred value of $x_B$ depends on the fitting window, and this dependence provides a useful diagnostic of the fit. If the fit begins too early, the residual is dominated by the larger pre-echo waveform and is only weakly sensitive to the onset of the secular contribution. If it begins too late, the turn-on is no longer resolved. Once the fitting interval lies entirely after the echo, a change in $x_B$ can be absorbed into a redefinition of the amplitude. Between these regimes, the inferred value of $2x_B$ develops a plateau as $t_{\min}$ is varied. Similarly, once $t_{\max}$ extends sufficiently far beyond the echo, the result becomes insensitive to further increases in the end time. This behavior is shown in Fig.~\ref{fig:xB_with_window}. Note that the minimum value of $2x_B$ used as a trial value is $2x_B\alpha_A=5$, so the preferred value for early and late $t_{\rm min}$ is railing against the bound, and not considered stable.

We propose using Eq.~\eqref{eq:free_freq_model_sec}, together with stability under variations of the fitting window, as a time-domain characterization of the spectral instability. The fitted $2x_B$ identifies the onset of the secular response, while $B_n/A_n$ characterizes its strength relative to the underlying mode. This provides a way to identify the instability directly from the waveform without assigning a time-independent shifted QNM frequency to the intermediate-time signal.

\section{Conclusion}

In this work we calculated the time-domain response of a wave equation with a double P\"oschl--Teller potential and identified the mechanism through which the spectral instability becomes visible in the waveform. The prompt ringdown retains the QNM frequencies of the unperturbed potential. Only after radiation has echoed, or interacted twice with the primary potential, does a term with a secular prefactor proportional to time appear. In the small-frequency-shift regime, this term can be interpreted as a perturbation of the original QNM frequencies, and its coefficient reproduces the analytic frequency-domain result of~\cite{Ianniccari:2024ysv}. Our numerical evolutions confirm both the causal onset of this effect and the predicted finite-time frequency behaviour. Even when the small-shift approximation breaks down, the prompt-plus-first-echo analytic waveform continues to describe the observed spectral migration during the first-echo window.

Previous work established that spectral instability does not generically translate into an immediate modification of the prompt ringdown, and that the contributions associated with the perturbed spectrum are causally delayed~\cite{Berti:2022xfj,Yang:2024vor}. Our results develop this picture by determining how the spectral migration proceeds once this delayed signal arrives. Rather than an instantaneous transition from the unperturbed to the perturbed QNM spectrum, the waveform passes through an intermediate regime governed by the secular response identified above. While this response reduces to the leading perturbative QNM frequency shift in the small-shift regime, more generally it gives rise to finite-time effective frequencies that need not coincide with either spectrum. This suggests that the intermediate-time behaviour of the spectral instability is better characterized by the onset and strength of the secular response than by a single shifted QNM frequency.

These results further emphasize the distinction between the global QNM spectrum and the causal time-domain response. A distant perturbation can substantially alter the global spectrum while the original QNM components retain their frequencies in the prompt response, within the separated-barrier approximation. This distinction is particularly relevant for extended environments around black holes, such as matter distributions outside the primary scattering region, whose effect on the ringdown need not be visible immediately even when their effect on the spectrum is large. At the same time, the delayed onset of the spectral instability shows that the full time-domain waveform can in principle provide information about the structure of the environment.

\section*{Acknowledgments}
AI disclosure: ChatGPT 6 Pro was used to proofread this paper. T.M.~would like to thank Romeo Felice Rosato for interesting discussions that prompted this work.
A.K.~thanks the Fundação para a Ciência e Tecnologia (FCT), Portugal, for the financial support to the FCT project ``Gravitational waves as a new probe of fundamental physics and astrophysics'' grant agreement 2023.07357.CEECIND/CP2830/CT0003.
N.F.~acknowledges funding from the FCT grant agreement 2023.06263.CEECIND/CP2830/CT0004.
The Center of Gravity is a Center of Excellence funded by the Danish National Research Foundation under grant No.~DNRF184.
We acknowledge support by VILLUM Foundation (grant no.~VIL37766).
V.C.~is a Villum Investigator.  
V.C.~acknowledges financial support provided under the European Union’s H2020 ERC Advanced Grant “Black holes: gravitational engines of discovery” grant agreement no.~Gravitas–101052587. 
Views and opinions expressed are however those of the author only and do not necessarily reflect those of the European Union or the European Research Council. Neither the European Union nor the granting authority can be held responsible for them.
We acknowledge FCT for the support to the Center for Astrophysics and Gravitation (CENTRA/IST/ULisboa) through FCT grant No.~UID/PRR/00099/2025 and grant No.~UID/00099/2025.
This project has received funding from the European Union's Horizon 2020 research and innovation programme under the Marie Sklodowska-Curie grant agreement No.~101007855 and No.~101131233.
This work is supported by Simons Foundation International~\cite{sfi} and the Simons Foundation~\cite{sf} through Simons Foundation grant SFI-MPS-BH-00012593-11.

\appendix

\section{List of Green's functions}
\label{App:list_greens_functions}

Here we give the Green's function for the P\"oschl-Teller potential \cite{Kuntz:2025gdq}, taking the cases of the main peak and the small bump separately, as described in the main text in Sec. \ref{sec:propagation_source_prelim}. 
The Green's function for a P\"oschl-Teller peak at (say) $x_B$, with the initial perturbation at $x_0$, and the observer at $x_1$, differs depending on the signs of $x_B-x_0$ and $x_B-x_1$. In other words, the Green's function depends on whether the source and the observer are to the left or the right of the potential peak. 
In the following we use $u = t-x$ and $v=t+x$, with subscripts to indicate the different points $u_i=t_i-x_i$. We list the Green's functions where the observer $x_1$ is to the right of the potential peak, $x_1>x_B$, in both cases where the source $x_0$ is situated to the left (transmission) or to the right (reflection) of the potential. To get the case with the observer to the left, a simple reflection around the peak can be performed: 
\begin{equation}
     G_B(x_1,x_0, t_1-t_0) =  G_B(2x_B-x_1,2x_B-x_0, t_1-t_0)
\end{equation}
and similarly for $G_A$, this equality being true because both of its sides solve the same equation with the same boundary conditions. The reflection formulas below assume that the observer lies farther from the peak than the source. The opposite ordering follows by exchanging the source and observer, using $G(x_1,x_0,t)=G(x_0,x_1,t)$. In either ordering, the causal condition is $t_1-t_0\geq|x_1-x_0|$. The expressions below depend on the QNM and Matsubara frequencies of both potentials $A$ and $B$, $\omega_n^A$, $\omega_n^B$, $w_k^A$ and $w_k^B$, defined in Sec.~\ref{sec:PT_potential}.

\subsubsection{Transmission through peak A}

The generic expression of the Green function in this case, $x_0<x_A=0<x_1$, is
\begin{equation}
    G_A(x_1,x_0, t_1- t_0) = \Theta(u_1-u_0) G_A^{\rm QNM}(x_1,x_0, t_1-t_0).
\end{equation}
We will always assume that $x_0$ and $x_1$ are situated far away from the potential. This gives
\begin{align}
    G_A^{\rm QNM} (x_1, & x_0, t_1-t_0) \approx\text{Re} \sum_{n \geq 0} \gamma_{n}^A e^{i \omega_n^A (u_0-u_1)} ,
\end{align} 
where
\begin{align}
    \gamma_n^A &= \frac{(-1)^{n+1}}{ n!} \frac{\Gamma \big( -2\lambda_A-n\big)}{\Gamma \big( \frac{1}{2} - \lambda_A - n \big)^2}.
\end{align}

\subsubsection{Reflection from peak A}

The Green function for $x_1\geq x_0>0$ is
\begin{align}
    G_A(x_1, x_0,& t_1- t_0) = \nonumber \\ &\Theta(u_1-u_0) \Theta(v_0-u_1) G_A^P(x_1,x_0, t_1-t_0)  \nonumber \\ & +\Theta(u_1-v_0) G_A^{\rm QNM}(x_1,x_0, t_1-t_0),
\end{align}
where the expression of the prompt and the QNM part of the Green function in the asymptotic regime where both $x_0$ and $x_1$ are far away to the right of the maximum of the potential is
\begin{align}
    G_A^P(x_1,x_0, t_1- t_0) &\simeq \sum_k \beta_k^A e^{i w_k^A(u_1-v_0)} \\
    G_A^{\rm QNM}(x_1,x_0, t_1-t_0) &\simeq \text{Re} \sum_{n \geq 0} (-1)^n \gamma_{n}^A e^{i \omega_n^A (v_0-u_1)}\ ,
\end{align}
with
\begin{equation}
    \beta_k^A = \frac{(-1)^{k+1}}{2 (k!)^2} \frac{\Gamma(1/2+\lambda_A+k) \Gamma(1/2-\lambda_A+k)}{\Gamma(1/2+\lambda_A) \Gamma(1/2-\lambda_A)}\ .
\end{equation}
The prompt sums run over $k=0,1,\ldots$, with $w_0^A=w_0^B=0$. The $k=0$ term gives the constant contribution $-1/2$.

\subsubsection{Transmission through peak B}

With respect to potential A, we have to take into account two differences: the potential is centered around $x=x_B$ and not around $x=0$. We are also interested in the case where $V_B$ is small, so that $4 V_B < \alpha_B^2$. The expressions for $\gamma_n^B$, and $\omega_n^B$ are the same as for potential A (Eq.~\eqref{eq:qnm_A} and Eq.~\eqref{eq:gammaA}). The only difference is that the $\lambda$ in the quasinormal mode frequency expression is given by Eq.~\eqref{eq:lambda_smallV}, as described in Sec. \ref{sec:PT_potential}. For this case, the quasinormal modes are purely imaginary. This changes the formulas in a few minor ways.
We have
\begin{align}
    G_B(x_1,x_0, &t_1-t_0) = \nonumber\\ &\Theta(u_1-u_0) G_B^{\rm QNM}(x_1,x_0, t_1-t_0)
\end{align}
with
\begin{align}
     G_B^{\rm QNM} (x_1,  x_0, &t_1-t_0) = \frac{1}{2} \sum_{n \geq 0} \gamma_{n}^B e^{i \omega_n^B (u_0-u_1)} \nonumber\\ &+ (\lambda_B \rightarrow - \lambda_B),
\end{align}

When we take the limit  $V_B/\alpha_B^2 \rightarrow 0$, we get that
\begin{equation}
     G_B^{\rm QNM} (x_1,  x_0, t_1-t_0) = - \frac{1}{2} e^{\frac{V_B}{\alpha_B}(u_0-u_1)}  + \mathcal{O}(V_B^2).
\end{equation}

\subsubsection{Reflection from peak B}

For $x_1\geq x_0>x_B$,
\begin{align}
    G_B(x_1,x_0, &t_1-t_0) = \nonumber\\ &\Theta(u_1-u_0) \Theta(v_0-u_1-2x_B) G_B^P(x_1,x_0, t_1-t_0)\nonumber \\ &+ \Theta(u_1-v_0+2x_B) G_B^{\rm QNM}(x_1,x_0, t_1-t_0),
\end{align}
where
\begin{align}
    G_B^P(x_1,x_0, t_1-t_0) &\simeq \sum_k \beta_k^B e^{i w_k^B(u_1-v_0+2x_B)}\; , \\
    G_B^{\rm QNM}(x_1,x_0, t_1-t_0) &\simeq \frac{1}{2}\sum_{n \geq 0} (-1)^n \gamma_{n}^B e^{i \omega_n^B (v_0-u_1-2x_B)}\nonumber\\ & + (\lambda_B \rightarrow - \lambda_B)\; .
\end{align}
Finally, considering the limit $V_B/\alpha_B^2 \rightarrow 0$ we find that, to linear order in $V_B/\alpha_B^2$,
\begin{align}
    G_B^P(x_1,x_0, &t_1-t_0) \simeq - \frac{1}{2} + \frac{V_B}{2 \alpha_B^2} \log \big[ 1 \nonumber \\ &+ \exp \alpha_B(u_1-v_0+2x_B) \big] \\
    G_B^{\rm QNM}(x_1,x_0,& t_1-t_0) \simeq - \frac{1}{2} \exp \bigg( \frac{V_B}{\alpha_B} (v_0 - u_1 - 2x_B) \bigg)\nonumber \\ & + \frac{V_B}{2 \alpha_B^2} \log \big( 1 + \exp \alpha_B(v_0-u_1-2x_B) \big)
\end{align}

\section{Details of the $n_{\rm echo}=1$ calculation}
\label{App:n=1calc}

In this appendix we provide the intermediate steps omitted from Sec.~\ref{sec:n=1}. In particular, we show explicitly the reflection from potential $B$ that leads to the expression for $\psi_2$ quoted in the main text, as well the reflection from potential $A$ that produces the resonant secular contribution in Eq.~\eqref{eq:psi3Resonant}. We work to leading order in $\epsilon_n=V_B/(\alpha_B\omega_n^A)$, with $|\epsilon_n|\ll1$.

\subsection{Reflection from potential $B$}

Starting from the expression for $\psi_2$ given in Sec.~\ref{sec:n=1},
\begin{equation}
\psi_2
=
-\int {\rm d}x_1
\left(
\partial_{x_1}G_B+\partial_{t_2}G_B
\right)\psi_1 ,
\end{equation}
we require the Green function for reflection from potential $B$. Since both $x_1$ and $x_2$ lie to the left of the barrier, its weak-barrier, large-separation form is
\begin{align}
G_B(x_2&,x_1,t_2-t_1)
={}\nonumber\\
&
\Theta(v_2-v_1)
\Theta(u_1-v_2+2x_B)
G_B^{\rm P}(x_2,x_1,t_2-t_1)
\nonumber\\
&+
\Theta(v_2-u_1-2x_B)
G_B^{\rm QNM}(x_2,x_1,t_2-t_1),
\label{eq:GBreflection_app}
\end{align}
where, using the Green functions listed in App.~\ref{App:list_greens_functions},
\begin{align}\label{eq:GBP_app}
G_B^{\rm P}
\simeq{}&
-\frac{1}{2}
+
\frac{V_B}{2\alpha_B^2}
\log\left[
1+e^{\alpha_B(v_2-u_1-2x_B)}
\right],
\end{align}
\begin{align}
    G_B^{\rm QNM}
    \simeq{}&
    -\frac{1}{2}
    e^{\frac{V_B}{\alpha_B}(u_1-v_2+2x_B)}
    \nonumber\\
    &+
    \frac{V_B}{2\alpha_B^2}
    \log\left[
        1+e^{\alpha_B(u_1-v_2+2x_B)}
    \right].
    \label{eq:GBQNM_app}
\end{align}

In the combination entering the convolution, the delta-function contributions arising from derivatives of the step functions cancel. The remaining derivatives are
\begin{align}
    \partial_{x_1}G_B^{\rm P}
    +\partial_{t_2}G_B^{\rm P}
    ={}&
    \frac{V_B}{\alpha_B}
    \frac{
        e^{\alpha_B(v_2-u_1-2x_B)}
    }{
        1+e^{\alpha_B(v_2-u_1-2x_B)}
    },
    \label{eq:GBPderiv_app}
\\
    \partial_{x_1}G_B^{\rm QNM}
    +\partial_{t_2}G_B^{\rm QNM}
    ={}&
    -\frac{V_B}{\alpha_B}
    \frac{
        e^{-\alpha_B(v_2-u_1-2x_B)}
    }{
        1+e^{-\alpha_B(v_2-u_1-2x_B)}
    }
    \nonumber\\
    &+
    \frac{V_B}{\alpha_B}
    e^{\frac{V_B}{\alpha_B}(u_1-v_2+2x_B)}.
    \label{eq:GBQNMderiv_app}
\end{align}

We are interested in the waveform after the reflected signal from $B$ has arrived, $v_2>u_0+2x_B$. The support of the Green function and of $\psi_1$ then splits the convolution into
\begin{align}
    \psi_2={}&
    -\Theta(v_2-u_0-2x_B)\nonumber\\&\times
    \int_{-\infty}^{t_1-v_2+2x_B}
    {\rm d}x_1\,
    \left(
        \partial_{x_1}G_B^{\rm P}
        +\partial_{t_2}G_B^{\rm P}
    \right)\psi_1
    \nonumber\\
    &-
    \Theta(v_2-u_0-2x_B)\nonumber\\&\times
    \int_{t_1-v_2+2x_B}^{t_1-u_0}
    {\rm d}x_1\,
    \left(
        \partial_{x_1}G_B^{\rm QNM}
        +\partial_{t_2}G_B^{\rm QNM}
    \right)\psi_1 .
    \label{eq:psi2split_app}
\end{align}

The first integral gives
\begin{equation}
    \left.\psi_2\right|_{\rm P}
    =
    \Theta(v_2-u_0-2x_B)
    \frac{V_B}{2\alpha_B^2}
    {\rm Re}\sum_n
    \gamma_n^A d_n
    e^{i\omega_n^A(u_0-v_2+2x_B)},
    \label{eq:psi2P_app}
\end{equation}
where
\begin{equation}
    d_n
    =
    D\left(
        \frac{1}{2}
        +\frac{i\omega_n^A}{2\alpha_B}
    \right)
    -
    D\left(
        1+\frac{i\omega_n^A}{2\alpha_B}
    \right),
    \label{eq:dn_app}
\end{equation}
and $D(z)\equiv {\rm d}\log\Gamma(z)/{\rm d}z$ is the digamma function.

The QNM part gives
\begin{align}
    \left.\psi_2  \right|_{\rm QNM}
    &=
    \Theta(v_2-u_0-2x_B)
    \frac{V_B}{\alpha_B}\nonumber\\&\times
    {\rm Re}\sum_n\gamma_n^A
    \Bigg\{
    e^{i\omega_n^A(u_0-v_2+2x_B)}
    \left(
        -\frac{\tilde d_n}{2\alpha_B}
        +\frac{1}{i\omega_n^A}
    \right)
    \nonumber\\
    &
    -\frac{1}{i\omega_n^A}
    e^{\frac{V_B}{\alpha_B}(u_0-v_2+2x_B)}
    \nonumber\\
    &
    +\frac{1}{i\omega_n^A}
    {}_2F_1\left(
        1,
        i\frac{\omega_n^A}{\alpha_B},
        1+i\frac{\omega_n^A}{\alpha_B},
        -e^{\alpha_B(v_2-u_0-2x_B)}
    \right)
    \Bigg\},
    \label{eq:psi2QNM_app}
\end{align}
with
\begin{equation}
    \tilde d_n
    =
    D\left(
        \frac{1}{2}
        +\frac{i\omega_n^A}{2\alpha_B}
    \right)
    -
    D\left(
        \frac{i\omega_n^A}{2\alpha_B}
    \right).
    \label{eq:dntilde_app}
\end{equation}

Adding Eqs.~\eqref{eq:psi2P_app} and \eqref{eq:psi2QNM_app}, we get that
\begin{align}
    \psi_2
    ={}&
    \Theta(v_2-u_0-2x_B)
    \frac{V_B}{\alpha_B}\nonumber\\&\times
    {\rm Re}\sum_n
    \frac{\gamma_n^A}{i\omega_n^A}
    \Bigg[
    e^{i\omega_n^A(u_0-v_2+2x_B)}
    \left(
        1+
        \frac{i\omega_n^A(d_n-\tilde d_n)}
             {2\alpha_B}
    \right)
    \nonumber\\
    &
    -
    e^{\frac{V_B}{\alpha_B}(u_0-v_2+2x_B)}\nonumber\\&
    +
    {}_2F_1\left(
        1,
        i\frac{\omega_n^A}{\alpha_B},
        1+i\frac{\omega_n^A}{\alpha_B},
        -e^{\alpha_B(v_2-u_0-2x_B)}
    \right)
    \Bigg].
    \label{eq:psi2combined_app}
\end{align}

The coefficient of the first term vanishes since $d_n-\tilde d_n=-\frac{2\alpha_B}{i\omega_n^A}$.
Using $\epsilon_n=V_B/(\alpha_B\omega_n^A)$, Eq.~\eqref{eq:psi2combined_app} therefore reduces to
\begin{align}
    \psi_2
    ={}&
    \Theta(v_2-u_0-2x_B)
    {\rm Re}\sum_n
    i\gamma_n^A\epsilon_n
    \Bigg[
    e^{\frac{V_B}{\alpha_B}(u_0-v_2+2x_B)}
    \nonumber\\
    &
    -
    {}_2F_1\left(
        1,
        i\frac{\omega_n^A}{\alpha_B},
        1+i\frac{\omega_n^A}{\alpha_B},
        -e^{\alpha_B(v_2-u_0-2x_B)}
    \right)
    \Bigg]\nonumber \\&
    +\mathcal{O}(\epsilon_n^2),
    \label{eq:psi2result_app}
\end{align}
which is the expression quoted in Sec.~\ref{sec:n=1}.

The late-time expansion of the hypergeometric function and the resulting part of $\psi_2$ oscillating at the QNM frequencies $\omega_n^A$ are given in the main text, culminating in Eq.~\eqref{eq:psi2QNMA}. We use that expression directly below.

\subsection{Resonant reflection from potential $A$}

We now evaluate the resonant part of the reflection from $A$. Starting from Eq.~\eqref{eq:psi3Green}, the QNM contribution to the required derivative of the reflected Green function is
\begin{equation}
    \partial_{x_2}G_A^{\rm QNM}
    -
    \partial_{t_3}G_A^{\rm QNM}
    =
    2\,{\rm Re}\sum_m
    (-1)^m
    \gamma_m^A i\omega_m^A
    e^{i\omega_m^A(v_2-u_3)}.
    \label{eq:GAderivative_app}
\end{equation}

Substituting Eq.~\eqref{eq:psi2QNMA} into Eq.~\eqref{eq:psi3Green}, the contribution containing the $A$-QNM frequencies is
\begin{align}
    \psi_3 \supset{}&
    \Theta(u_3-u_0-2x_B)
    \int_{u_0+2x_B-t_2}^{u_3-t_2}
    {\rm d}x_2
    \nonumber\\
    &\times
    \left[
    {\rm Re}\sum_n
    \gamma_n^A\epsilon_n
    e^{i\omega_n^A(u_0-v_2+2x_B)}
    \frac{
        \pi\omega_n^A/\alpha_B
    }{
        \sin(\pi i\omega_n^A/\alpha_B)
    }
    \right]
    \nonumber\\
    &\times
    \left[
    2\,{\rm Re}\sum_m
    (-1)^m
    \gamma_m^A i\omega_m^A
    e^{i\omega_m^A(v_2-u_3)}
    \right].
\label{eq:psi3double_app}
\end{align}

For $m\neq n$, the integrand retains an $x_2$-dependent phase through $v_2=t_2+x_2$, and integration produces a sum of the same quasinormal mode frequency terms, only with changed amplitudes. However, for the resonant terms $m=n$, the $v_2$ dependence cancels. The integrand is then independent of $x_2$, and the integral gives
\begin{equation}
    \int_{u_0+2x_B-t_2}^{u_3-t_2}{\rm d}x_2
    =
    u_3-u_0-2x_B.
\end{equation}
The resonant contribution is therefore
\begin{align}
    \psi_3 \supset{}&
    (u_3-u_0-2x_B)
    \Theta(u_3-u_0-2x_B)\nonumber\\&\times
    {\rm Re}\sum_n
    (-1)^n
    \omega_n^A
    (\gamma_n^A)^2
    \epsilon_n
    \nonumber\\
    &\times
    e^{i\omega_n^A(u_0-u_3+2x_B)}
    \left[
        \frac{
            \pi i\omega_n^A/\alpha_B
        }{
            \sin(\pi i\omega_n^A/\alpha_B)
        }
    \right]
    +\mathcal{O}(\epsilon_n^2),
    \label{eq:psi3res_app}
\end{align}
which reproduces Eq.~\eqref{eq:psi3Resonant}.

Finally, the term in Eq.~\eqref{eq:psi3res_app} is already of order $\epsilon_n$. As was shown in Sec.~\ref{sec:n=0}, transmission through the weak potential $B$ changes the amplitude of an $A$-QNM component only by a relative correction of order $\epsilon_n$, and introduces $B$-mode contributions at the same relative order. These effects therefore enter the first-echo waveform only at $\mathcal{O}(\epsilon_n^2)$ and are consistently neglected here. The resonant contribution to the observer waveform is then Eq.~\eqref{eq:psi3res_app}, with $u_3$ replaced by $u_{\rm obs}$.

\section{Numerical study}
\label{App:numerics}

\subsection{Initial Data}
\label{App:initial_data}

In order to obtain the ringdown signal in the main text we construct initial data that closely resembles the first overtone of the unperturbed potential close to the origin. We use the analytic solution for the overtone of the P\"oschl--Teller potential from \cite{Berti:2009kk}. We multiply that solution by a window function to suppress the blow up as $|x|\rightarrow\infty$ in standard coordinates. The window function we choose is given by
\begin{align}
    W_n(x)=
\frac{1}{1+\exp\left(
\frac{(2n+1)(|x|-r_{\rm cut})}{w_{\rm width}}
\right)},
\label{eq:window_function}
\end{align}
where $n$ is the overtone number. For the fundamental mode with $V_A=1$, $\alpha_A=1$, we find that $w_{\rm width} = 0.5$ and $r_{\rm cut}=10$ works well and excites a superposition of the prograde and retrograde $n=0$ modes. For the first overtone, we find that $w_{\rm width} = 0.8$ and $r_{\rm cut}=10$ gives a strong excitation of the first overtone, with no fundamental mode resolved before the precision floor is reached in the unperturbed evolution.

\begin{figure}[h]
\centering
\includegraphics[width=8cm]{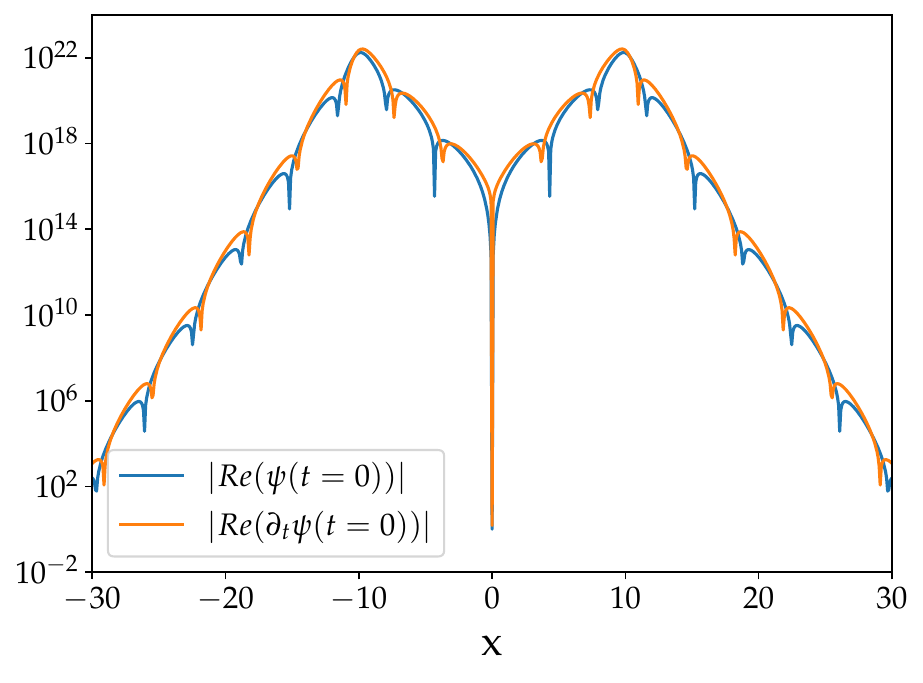}
\caption{The initial data used for overtone dominated evolution. A window function is used to control the exponential growth with $x$ at larger distances.
}
\label{fig:ID_profile}
\end{figure}

The profile of our initial data then closely resembles an overtone close to the origin (where the potential is at its maximum), and far from the origin has the same real frequency as the overtone but is decaying with distance, instead of growing. The profile of the initial data for the case considered in the main text is plotted in Fig.~\ref{fig:ID_profile}.

One might worry that the initial recovery of the unperturbed overtone frequency in Fig.~\ref{fig:free_freq_fit} is just the direct propagation of the initial data to infinity. We use $t_{\rm ring}\simeq x_{\rm obs}+r_{\rm cut}$ as a reference for the onset of clean ringdown in these evolutions. Since the window decays smoothly, this is not an exact boundary of the signal's causal support. The window is not compactly supported, and $r_{\rm cut}>x_B$ in the example studied here, so the numerical initial data extend across the secondary barrier. The fits in Sec.~\ref{sec:confirmation_numerics_timedomain} test whether the departure from the unperturbed overtone occurs after the expected round-trip delay relative to this reference time.

\begin{figure}[h]
\centering
\includegraphics[width=8cm]{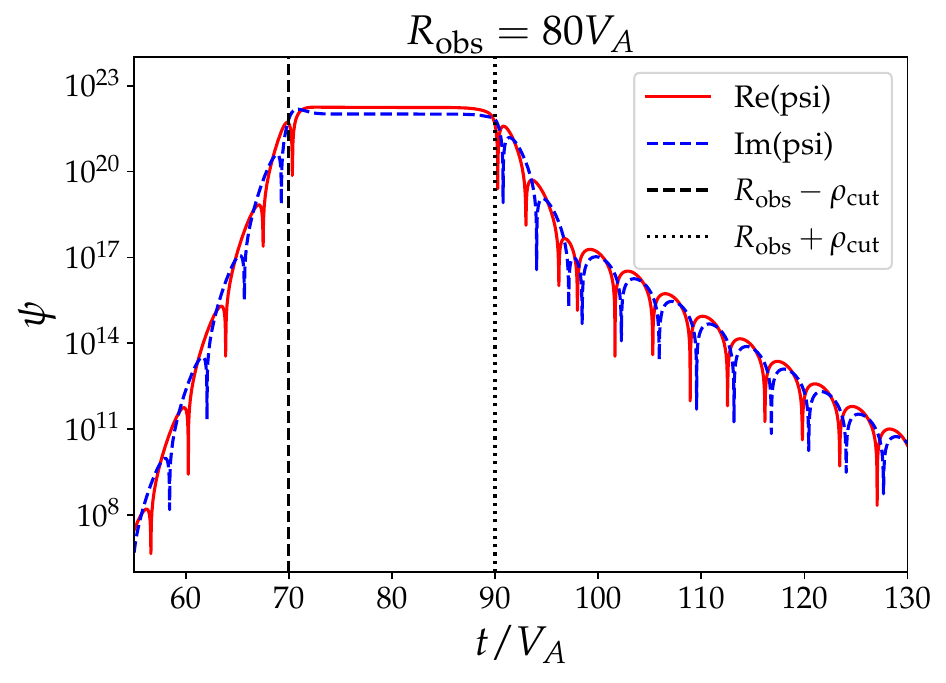}
\caption{The plateau begins near $t=x_{\rm obs}-r_{\rm cut}$, when radiation from the window transition on the right reaches the observer. It contains the feature produced by this transition together with the rest of the initial data travelling directly to the observer.
The time $t\simeq x_{\rm obs}+r_{\rm cut}$ marks the arrival of radiation from the window transition on the left after it has interacted with the primary peak.} 
\label{fig:full_waveform}
\end{figure}

The full waveform, including the initial data that propagate directly to the observer, is shown in Fig.~\ref{fig:full_waveform}. The clean ringdown begins near the time when radiation from the window transition at $x=-r_{\rm cut}$ has reached the primary peak and then the observer. We check this timing by varying $r_{\rm cut}$ while keeping the rest of the initial data fixed. For the initial data and parameter choices studied here, the relative fundamental-mode amplitude is approximately proportional to $V_B/V_A$. Its slower decay eventually limits the interval over which the overtone can be fitted reliably.

\subsection{Convergence}
\label{App:convergence}

\begin{figure}[h]
\centering
\includegraphics[width=8cm]{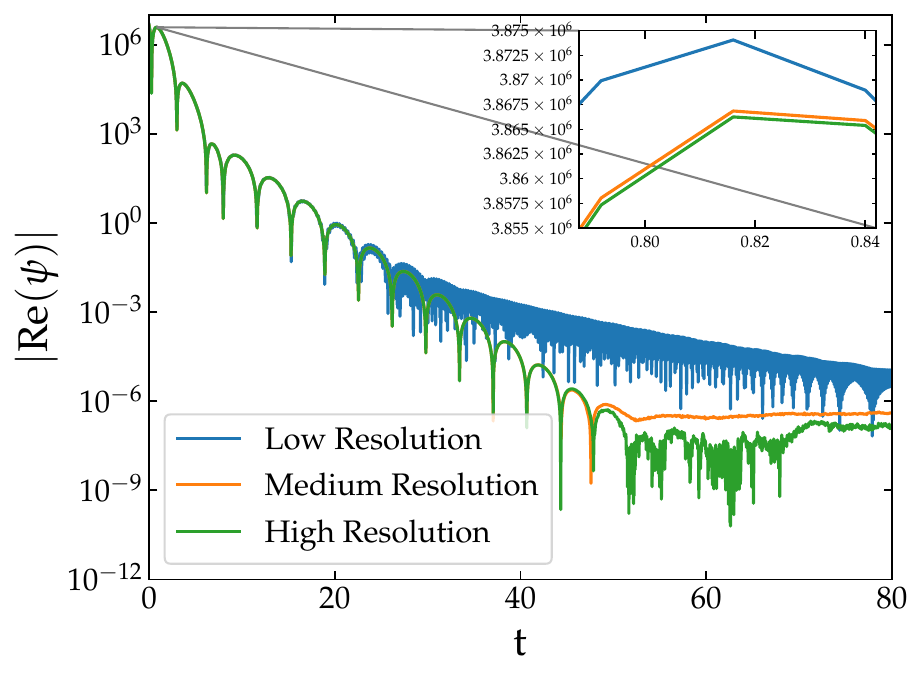}
\caption{The ringdown component of simulated data for three different resolutions. Here the potential is the same as that used for the results in Fig. \ref{fig:free_freq_fit}, and the medium resolution is the resolution used for that result. The signal at early times is dominated by the first overtone due to our choice of initial data as is described in Appendix \ref{App:initial_data}.
}
\label{fig:res_comparison}
\end{figure}

We use a fourth order finite difference code to generate the ringdown signal. In Fig. \ref{fig:res_comparison}, we show the signal using low, medium, and high resolution runs where each increase in resolution is by a factor of two. Running a free frequency fit on the medium and high resolutions, as performed in Fig. \ref{fig:free_freq_fit}, results in frequencies that only differ noticeably at very late times. For the time range shown in Fig. \ref{fig:free_freq_fit} the results are indistinguishable at the scale shown. 

\section{Frequency-domain computation of QNMs}\label{App:freq}

The calculation of the QNM frequencies in the frequency domain follows closely the method of~\cite{Jaramillo:2020tuu}. A transformation to hyperboloidal coordinates $\{\tau,\sigma\}$ is used, such that $t=\tau-h(\sigma)$ and $x=g(\sigma)$. $\sigma$ is a compactified spatial coordinate, such that the domain $x\in(-\infty,\infty)$ is mapped onto $\sigma\in(-1,1)$. The height function $h(\sigma)$ then satisfies $h\sim g$ in the $\sigma\to -1$ limit, and $h\sim -g$ in the $\sigma\to 1$ limit. To begin with, we performed the computation with the same choice of functions as in~\cite{Jaramillo:2020tuu}, namely
\begin{align}
	h(\sigma)&=\frac{1}{2}\log(1-\sigma^2),\\
	g(\sigma)&=\operatorname{arctan}(\sigma),\label{x_compact}
\end{align}
for which the wave equation takes on the form
\begin{equation}\label{wave2}
	-\partial^2_\tau\psi+L_1\psi+L_2\partial_\tau\psi=0,
\end{equation}
where the spatial differential operators are
\begin{equation}
	\begin{split}
		L_1&=\frac{p}{w}\partial_\sigma^2+\frac{p'}{w}\partial_\sigma-\frac{q}{w},\\
		L_2&=2\frac{\gamma}{w}\partial_\sigma+\frac{\gamma'}{w},
	\end{split}
\end{equation}
with
\begin{equation}\label{op_functions}
	\begin{split}
		w&=1,\quad p=1-\sigma^2,\quad \gamma=-\sigma,\\
		q&=V_A+\frac{V_B}{(1-\sigma^2)\cosh^2\left(\alpha_B(\text{arctanh}(\sigma)-x_B)\right)},
	\end{split}
\end{equation}
in units of $\alpha_A=1$. The domain $\sigma\in[-1,1]$ (including the boundaries, which usually correspond to horizons and asymptotic regions) was then discretised into a Chebyshev-Lobatto grid, and the spatial differential operators were discretised using a pseudo-spectral method, as outlined in~\cite{Jaramillo:2020tuu}, and the QNM spectrum was subsequently obtained through an eigenvalue calculation.

We note that for the choice of spatial compactification function \eqref{x_compact}, the secondary Pöschl-Teller peak is very poorly resolved in the discretised $\sigma$ grid for values of $x_B\gtrsim5/\alpha_A$, as this peak would be too close to the boundary where the compactification diverges. To resolve this issue, we tested several alternative choices of compactification functions, finally choosing $g$, and its corresponding $h$, to be
\begin{align}
    h(\sigma)&=\frac{3}{8}(1-x_B)\sigma^2(1-\sigma^2)+\frac{1}{2}\log(1-\sigma^2),\\
    g(\sigma)&=-\frac{1}{2}(1-x_B)\sigma(3-\sigma^2)-\frac{1}{2}\log\left(\frac{1-\sigma}{1+\sigma}\right).\label{x_compact2}
\end{align}
This in turn changes slightly the expressions for the functions in (D5), the details of which we omit for brevity. For the separation used in the example in this work, $x_B=4/\alpha_A$, the choice \eqref{x_compact} already gives convergent results for the first $5$ overtones with $N=150$ gridpoints, while using \eqref{x_compact2} gives 7 overtones for the same resolution. The difference becomes much more drastic for larger values of $x_B$, where the choice \eqref{x_compact} quickly becomes unusable for resolutions $N\sim10^2$, while \eqref{x_compact2} still gives convergent results.

\bibliography{bib}

\end{document}